\documentclass[a4paper,11pt]{article}
\usepackage{jheppub} % for details on the use of the package, please
\usepackage{enumerate}
\usepackage{latexsym}
\usepackage{makeidx}
\usepackage{tikz}
\usepackage{hyperref}
\usepackage{amsfonts}
\usepackage{amsmath}
\usepackage{amssymb}
\usepackage{slashed}
\usepackage{graphicx,graphics,color}
\usepackage{multirow}
\usepackage{upgreek}
\def\be{\begin{equation}}
	\def\bea{\begin{eqnarray}}
	\def\ee{\end{equation}}
	\def\eea{\end{eqnarray}}

\def\nn{\nonumber \\}
\def\p{\partial}

\def\D{\operatorname{Ad}}
\def\on{\operatorname}
\def\({\left(}
\def\){\right)}
\def\[{\left[}
\def\]{\right]}

\def\p{\partial}

\def \mcal{\mathcal}
\def \la{\langle}
\def \ra{\rangle}

\title{\boldmath Deformed Integrable Models from Holomorphic Chern-Simons Theory}

 \author{Bin Chen$^{1,2,3}$, Yi-jun He$^1$, and Jia Tian$^4$}
\affiliation{$^{1}$School of Physics, Peking University, No.5 Yiheyuan Rd, Beijing 100871, P.~R.~China\\
		$^{2}$Collaborative Innovation Center of Quantum Matter, No.5 Yiheyuan Rd, Beijing 100871, P.~R.~China\\
		$^{3}$Center for High Energy Physics, Peking University, No.5 Yiheyuan Rd, Beijing 100871, P.~R.~China\\
		$^4$Kavli Institute for Theoretical Sciences (KITS),
University of Chinese Academy of Science, 100190 Beijing, P. R. China}

\emailAdd{bchen01@pku.edu.cn,yjhe96@pku.edu.cn,wukongjiaozi@ucas.ac.cn}

\abstract{We study the approaches to two-dimensional integrable field theories via a six-dimensional(6D) holomorphic Chern-Simons theory defined on twistor space. Under symmetry reduction, it reduces to a four-dimensional Chern-Simons theory, while under solving along fibres it leads to four-dimensional(4D) integrable theory, the anti-self-dual Yang-Mills or its generalizations. From both four-dimensional theories, various two-dimensional integrable field theories can be obtained. In this work, we try to investigate several two-dimensional integrable deformations in this framework. We find that the $\lambda$--deformation, the rational $\eta$--deformation and the generalized $\lambda$--deformation can not be realized from 4D integrable  model approach, even though they could be obtained from 4D Chern-Simons theory. The obstacle stems from the incompatibility between the symmetry reduction and the boundary conditions. Nevertheless, we show that a coupled theory of $\lambda$-deformation and  the $\eta$-deformation in the trigonometric description could be obtained from the 6D theory in both ways, by considering the case that $(3,0)$-form in the 6D theory is allowed to have zeros. }

\begin{document} 
\maketitle
\flushbottom

\section{Introduction}

One interesting issue on integrable systems is to find a unifying way to organize various kinds of integrable models. It was found \cite{Twistor} that many two-dimensional(2D) integrable models could be organized by four dimensional anti-self-dual Yang-Mills (ASDYM) equations via symmetry reductions, by viewing the Lax equation as a zero curvature condition.
  By the Penrose-Ward transformation, the ASDYM connections are identified with  holomorphic vector bundles on the twistor space \cite{penrose1984spinors,ward1990twistor} and the spectral parameter of an integrable system is interpreted as a coordinate on the twistor bundle from this geometric point of view. Quite recently, another geometric perspective on two dimensional integrable field theories has been proposed in \cite{CWY}.  In this new approach, the two dimensional integrable field theories can be systematically constructed by inserting two dimensional defects into a novel four dimensional(4D) Chern-Simons (4DCS) theory\cite{Costello:2013zra} which is specified by a holomorphic 1-form $\omega$, generalizing the study on lattice integrable models in \cite{Witten:2016spx,Costello:2017dso,Costello:2018gyb}. This 4D gauge theory approach has attracted much attention and has been under intense study since its proposal \cite{Delduc:2019whp,Bassi:2019aaf,Fukushima:2020kta,Fukushima:2020dcp,Costello:2020lpi,Tian:2020ryu,Tian:2020meg,Benini:2020skc,Skiner,Penna:2020uky,Lacroix:2020flf,Caudrelier:2020xtn,Fukushima:2020tqv}. 

The relationship between these two geometric approaches was investigated in \cite{Costello,Skiner} from the perspective of a six dimensional holomorphic Chern-Simons theory (6DhCS) which is specified by a meromorphic $(3,0)$-form $\Omega$.  Starting from 6DhCS theory, one can either perform symmetry reduction on the twistor space to get the 4DCS, or insert defects, dubbed as solving along fibres in \cite{Skiner}, to obtain a four dimensional integrable theory which is a generalization of ASDYM. An essential feature is that appropriate boundary conditions on the fields have to be imposed at the locations of the poles of $\omega$ and $\Omega$ to ensure the gauge invariance. After imposing the boundary conditions, the field equations can be solved uniquely up to a gauge transformation.  Schematically the relationship is illustrated in the diagram showing in Fig. 1:
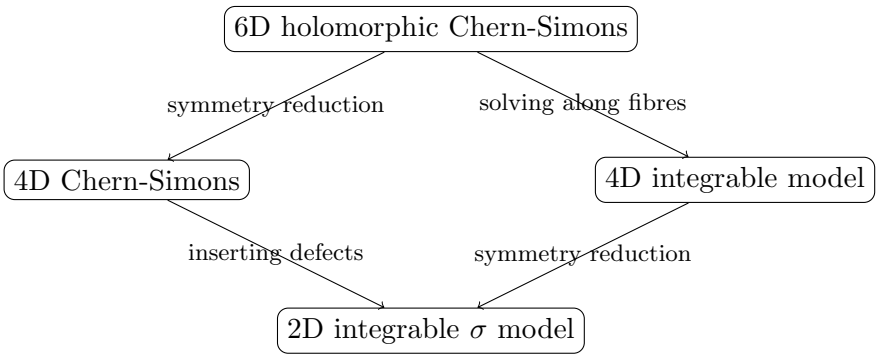
\begin{figure}[h]
\label{Figure1}
\begin{center}
\begin{tikzpicture}[scale=1]
\path (0,0) node[rectangle,
           rounded corners,draw](2D)  {2D integrable $\sigma$ model}
(0,4) node[rectangle,
           rounded corners,draw] (6D) {6D holomorphic Chern-Simons}
(-4,2) node[rectangle,
           rounded corners,draw] (4DCS) {4D Chern-Simons}
(4,2) node[rectangle,
           rounded corners,draw] (4DI) {4D integrable model};
\draw[->] (6D) -- (4DCS)
 node[pos=0.5]{{\footnotesize symmetry reduction}};
\draw[->] (6D) -- (4DI)
node[pos=0.5]{{\footnotesize solving along fibres}};
\draw[->] (4DCS) -- (2D)
node[pos=0.5]{{\footnotesize inserting defects}};
\draw[->] (4DI) -- (2D)
node[pos=0.5]{{\footnotesize symmetry reduction}};
\end{tikzpicture}
\end{center}
\caption{Relationship between 4DCS and ASDYM.}
\end{figure}

In \cite{Skiner}, the authors focused on the simplest example, the 2D principal chiral model\footnote{The possible trigonometric deformation and a coupled $\sigma$--model were also considered in \cite{Skiner}.}, in which case the resulting 4D integrable model is a WZW model \cite{4DWZW}(the Chalmers-Siegel action for ASDYM \cite{CSA}). In this case, the above diagram turns out to be commutative. It would be interesting to explore if other 2D integrable models can be fitted into the above diagram, as suggested in \cite{Skiner}. 

In this work, we would like to pick two important classes of deformed integrable models and consider their constructions from 6D holomorphic Chern-Simons theory. From 4DCS theory, 
various integrable deformations of 2D integrable models have been  constructed  by inserting more general defects which are described by Manin triples\cite{Bassi:2019aaf,Fukushima:2020kta,Fukushima:2020dcp,Tian:2020ryu,Tian:2020meg}. Among them two important examples are the $\lambda$-deformed principal chiral model \cite{Lambda} and the Yang-Baxter  $\sigma$-model \cite{YB1,YB2}. One compelling question is to investigate if these models can be read from 4D ASDYM and more generally what the diagram in Fig. \ref{Figure1} would look like when more general defects are inserted into 6DhCS.
In this paper, we show that when we consider more general defects the diagram in Fig. \ref{Figure1} is not completely commutative. This loss of commutativity originates from the fact that the symmetry reduction process may not be compatible with the boundary conditions. The symmetry reduction from 6DhCS to 4DCS suggests the \textit{matching conditions} between the 6D gauge connection and 4D gauge connection. Using this \textit{matching conditions}, we can 
read the boundary conditions in one theory from the ones in the other. However, the induced boundary conditions are often problematic in the sense that they cannot remove the gauge freedoms. On the other hand, even if we discard the \textit{matching conditions} and consider appropriate boundary conditions to remove gauge freedoms, it is still hard to find 
 interesting 4D integrable deformations, whose symmetry reduction would give rise to the 2D deformed model we want. 
In particular, it is impossible to construct a non-trivial 4D  $\lambda$--deformed WZW model from the 6DhCS. We find that the resulting 2D model is either undeformed or with the deformation parameter being restricted to specific values. This is because the $(3,0)$-form $\Omega$ in 6DhCS is too restrictive to allow non-trivial defect.  Therefore, we consider the situation that  $\Omega$ is of zeros such that non-trivial defects are allowed, and we discard the \textit{matching conditions} at the same time. We will investigate two cases, one being that $\Omega$ is of zeros and a fourth-order pole, the other being that $\Omega$ is of zeros and two double poles. The former case leads to 2D $\lambda$-deformed model coupled with an additional field, while the latter one leads to  the trigonometric description of the Yang-Baxter deformation \cite{Trig1,Trig2}. In both cases, we can generate a diagram like Fig. \ref{Figure1}.% for the Yang-Baxter model, after discarding the \textit{matching conditions}.

This paper is organized as follows. In section 2 we review the holomorphic Chern-Simons theory on the twistor space and the constructions in Fig. \ref{Figure1} described in \cite{Skiner} with a particular emphasis on the \textit{matching conditions} and the commutativity of the diagram. In section 3 we consider the $\lambda$-deformation and show a direct lift of the boundary condition associated with the poles of $\omega$ in 4DCS to the boundary condition associated with the poles of $\Omega$ in 6DhCS through the \textit{matching conditions} is problematic. In section 4 we extend our analysis to other possible defects and show that in general they do not lead to desired deformations. In section 5, we consider the case that $\Omega$ is of zeros and a fourth-order pole, and obtain a coupled $\lambda$-deformation.  
In section 6 we show that the Yang-Baxter deformation in a trigonometric description can actually be constructed in both ways. In this case both 4D and 2D version of deformed theory are obtained but with a necessary violation of the \textit{matching condition}. 

\section{Integrable field theories from holomorphic Chern-Simons theory}
\begin{figure}[h]
\label{Figure2}
\begin{center}
\begin{tikzpicture}[scale=1]
\path (0,0) node[rectangle,
           rounded corners,draw](2D)  {2D Principal Chiral model with WZW term}
(0,4) node[rectangle,
           rounded corners,draw] (6D) {6D holomorphic Chern-Simons}
(-4,2) node[rectangle,
           rounded corners,draw] (4DCS) {4D Chern-Simons}
(4,2) node[rectangle,
           rounded corners,draw] (4DI) {4D WZW model};
\draw[->] (6D) -- (4DCS)
 node[pos=0.5]{{\footnotesize symmetry reduction}};
\draw[->] (6D) -- (4DI)
node[pos=0.5]{{\footnotesize solving along fibres}};
\draw[->] (4DCS) -- (2D)
node[pos=0.5]{{\footnotesize inserting defects}};
\draw[->] (4DI) -- (2D)
node[pos=0.5]{{\footnotesize symmetry reduction}};
\end{tikzpicture}
\caption{A guide to the constructions of 4D WZW and 2D principle chiral model.}
\end{center}
\end{figure}
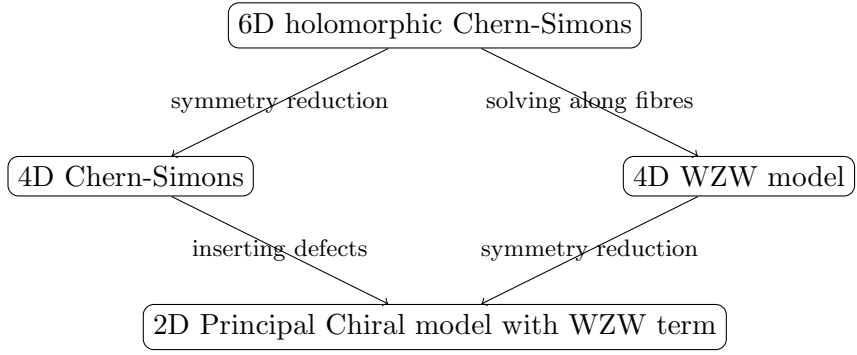

In this section we will review the constructions\footnote{We will adopt the convention used in \cite{Skiner}.} depicted by the diagram in Fig. \ref{Figure2}. 

\subsection{Holomorphic Chern-Simons on twistor space}
The starting point is the holomorphic Chern-Simons theory on the twistor space. The action of the theory is of the form
\begin{align}
    \label{equ:1.1}
    S_{\Omega}[\bar{\mathcal{A}}]=\frac{1}{2 \pi i} \int_{\mathbb{PT}} \Omega \wedge \mathrm{hCS}(\bar{\mathcal{A}}),
\end{align}
where $\mathrm{hCS}(\bar{\mathcal{A}})$ is of the form of Chern-Simons action
\begin{align}
    \label{equ:1.2}
    \mathrm{hCS}(\bar{\mathcal{A}})=\operatorname{tr}\left(\bar{\mathcal{A}} \wedge \bar{\partial} \bar{\mathcal{A}}+\frac{2}{3} \bar{\mathcal{A}} \wedge \bar{\mathcal{A}} \wedge \bar{\mathcal{A}}\right),
\end{align}
and $\Omega$ is a meromorphic $(3,0)$-form which may contain both zeros and poles.  The $(3,0)$-form $\Omega$ can be rewritten as 
\begin{align}
    \label{add:1.1}
    \Omega=\mathrm{D}^3 Z\otimes \Phi,
\end{align}
where 
\begin{align}
    \label{add:1.2}
    \mathrm{D}^3 Z=\frac{\langle\mathrm{d}\pi \pi\rangle\wedge\mathrm{d}^2 x^{A'B'}\pi_{A'}\pi_{B'}}{2},
\end{align}
is the canonical holomorphic 3-form on $\mathbb{CP}^3$ in the spinor notation and  $\Phi$ which encodes the analytic properties of $\Omega$ is a meromorphic section of $\mathcal{O}(-4)\rightarrow\mathbb{PT}$. We restrict to the case in which $\Phi$ only depends on the fibre direction of twistor space $\mathbb{PT}\rightarrow\mathbb{CP}^1$, i.e. $\Phi$ depends only on the coordinates $\pi^{A'}$. To derive the principal chiral model, we should take  $\Phi=(\langle\pi\alpha\rangle\langle\pi\beta\rangle)^{-2}$ which was first proposed by Costello in a seminar \cite{Costello} and was concretely realized by Bittleston and Skinner in \cite{Skiner}. 

The partial gauge connection $\bar{\mathcal{A}}$ in \eqref{equ:1.2} can be decomposed in terms of basis of $(0,1)$-forms 
\begin{align}
\label{equ:1.3}
    \bar{\mathcal{A}}=\bar{e}^0\bar{\mathcal{A}}_0+\hat{e}^{A}\hat{\mathcal{A}}_{A}, 
\end{align}
where $\bar{e}^0$ is in the direction of $\mathbb{CP}^1$ fibre over $\mathbb{E}^4$ and $\hat{e}^{A}$ are in the directions of $\mathbb{E}^4$.  
At the positions of  the poles of $\Omega$, the variation of the action $S_{\Omega}[\bar{\mathcal{A}}]$ generates the boundary terms 
\begin{align}
\label{add:1.5}
    \delta S_{\Omega}|_{bdy}
    =&\frac{1}{2 \pi i} \lim_{\epsilon \rightarrow 0} \int_{\mathbb{P T}_{\epsilon}} \mathrm{d}(\Omega \wedge \operatorname{tr}(\delta \bar{\mathcal{A}} \wedge \bar{\mathcal{A}}))\notag\\
    =&\frac{1}{4 \pi i} \sum_{z\in\mathfrak{p}} \lim _{\epsilon \rightarrow 0} \oint_{S_{z, \epsilon}^{1}}\left[\langle\mathrm{d} \pi \pi\rangle\Phi \int_{\mathbb{E}^{4}} \mathrm{~d}^{2} x^{A^{\prime} B^{\prime}} \pi_{A^{\prime}} \pi_{B^{\prime}} \wedge \operatorname{tr}(\delta \bar{\mathcal{A}} \wedge \bar{\mathcal{A}})\right],
\end{align}
where $\mathfrak{p}$ is the set of the poles of $\Omega$ and $S_{z, \epsilon}^{1}$ is a circle of radius $\epsilon$ around the poles $\pi=z$. Using the following identity
\begin{align}
    \label{equ:1.6}
    \mathrm{d}^{2} x^{A^{\prime} B^{\prime}} \pi_{A^{\prime}} \pi_{B^{\prime}} \wedge \hat{e}^{A} \wedge \hat{e}^{B}=-2 \varepsilon^{A B} \mathrm{vol}_{4},
\end{align}
we find that
\begin{align}
\label{add:1.7}
    \delta S_{\Omega}|_{bdy}
    =&-\frac{1}{2 \pi i} \sum_{z\in\mathfrak{p}} \lim _{\epsilon \rightarrow 0} \oint_{S_{z, \epsilon}^{1}}\left[\langle\mathrm{d} \pi \pi\rangle\Phi \int_{\mathbb{E}^{4}} \mathrm{vol}_4 \varepsilon^{AB} \operatorname{tr}(\delta \bar{\mathcal{A}_A} \bar{\mathcal{A}}_B)\right],
\end{align}
which should be set to $0$ by choosing appropriate boundary conditions in order to make the variation well defined.

\subsection{Symmetry reduction to 4D Chern-Simons theory}
Firstly we follow the left route  in Fig. \ref{Figure2}. It was shown in \cite{Skiner} that the 4DCS theory can be obtained by performing a symmetry reduction by a 2-dimensional group of translations $H$ in the  holomorphic Chern-Simons theory. The generators of $H$ are chosen to be the translations along the complex null vectors
\begin{align}
    \label{equ:1.8}
    \chi=\hat{\kappa}^{A} \hat{\mu}^{A^{\prime}} \partial_{A A^{\prime}}=\partial_{z}, \qquad \bar{\chi}=\kappa^{A} \mu^{A^{\prime}} \partial_{A A^{\prime}}=\partial_{\bar{z}},
\end{align}
 where we have introduced double-null complex coordinates on the 4D Euclidean space $\mathbb{E}^4$ as\footnote{Here we take $\|\kappa\|^{2}=\|\mu\|^{2}=1$. }
\begin{align}
    \label{equ:1.9}
    z=x^{A A^{\prime}} \kappa_{A} \mu_{A^{\prime}}, \quad w=-x^{A A^{\prime}} \hat{\kappa}_{A} \mu_{A^{\prime}}, \quad \bar{w}=x^{A A^{\prime}} \kappa_{A} \hat{\mu}_{A^{\prime}}, \quad \bar{z}=x^{A A^{\prime}} \hat{\kappa}_{A} \hat{\mu}_{A^{\prime}}. 
\end{align}
For future convenience we also define 
\begin{align}
    \omega=\kappa^A\hat{\mu}^{A'}\partial_{AA'}=\partial_w, \qquad \bar{\omega}=-\hat{\kappa}^A\mu^{A'}\partial_{AA'}=\partial_{\bar{w}}. \notag
\end{align}
 The invariant condition of the connection is
\begin{align}
\label{equ:1.10}
    \mathcal{L}_{X}(\mathrm{D}s)=\mathrm{D}(\mathcal{L}_{X}s), \qquad X\in\mathfrak{h},
\end{align}
where $s\in\Gamma(\mathcal{E})$ and $\mathcal{E}$ is a vector bundle over an open subset $\mathcal{U}$ of the twistor space $\mathbb{PT}$ and the symbol $\mathcal{L}_{X}$ denotes the Lie derivative which is defined by  lifting the action of $H$ on $\mathcal{U}$ to $\mathcal{E}$. 
The action of translation group $H$ is necessarily free such that it is always possible to find an invariant gauge where the invariant condition \eqref{equ:1.10} becomes
\begin{align}
    \label{equ:1.11}
    \mathcal{L}'_{\chi} \bar{\mathcal{A}}=\mathcal{L}'_{\bar{\chi}} \bar{\mathcal{A}}=0,
\end{align}
where $\mathcal{L}'$ denotes the ordinary Lie derivative operator acting on differential forms. In this gauge choice, the residual gauge freedom consists of gauge transformations generated by $\chi$ and $\bar\chi$. 

The Lagrangian density of the holomorphic Chern-Simons theory is invariant under the action of $H$ as well so one is able to do a dimensional reduction by integrating out $H$ directions. However it is more convenient to do this reduction  by contracting the bivector $\chi\wedge\bar{\chi}$ with the Lagrangian density of the holomorphic Chern-Simons theory. 
Performing this contraction directly to \eqref{equ:1.2} and \eqref{add:1.1} gives
\begin{align}
    \iota_{\bar\chi}\Omega=&\langle\mathrm{d}\pi \pi\rangle\wedge(\mathrm{d}\bar{w}\langle\mu\pi\rangle-\mathrm{d}z\langle\hat{\mu}\pi\rangle)\langle\mu\pi\rangle\Phi,\notag\\
    \iota_{\chi}\Omega=&\langle\mathrm{d}\pi \pi\rangle\wedge(\mathrm{d}\bar{z}\langle\mu\pi\rangle+\mathrm{d}w\langle\hat{\mu}\pi\rangle)\langle\hat{\mu}\pi\rangle\Phi,\notag\\
    \iota_{\chi\wedge\bar{\chi}}\Omega=&\langle\mathrm{d}\pi \pi\rangle\langle\pi\mu\rangle\langle\pi\hat{\mu}\rangle\Phi,\notag
\end{align}
and
\begin{align}
    \iota_{\bar\chi}\operatorname{hCS}(\bar{\mathcal{A}})=&2\operatorname{tr}\left((\iota_{\bar\chi}\bar{\mathcal{A}})\bar{\mathcal{F}}\right),\notag\\
    \iota_{\chi}\operatorname{hCS}(\bar{\mathcal{A}})=&2\operatorname{tr}\left((\iota_{\chi}\bar{\mathcal{A}})\bar{\mathcal{F}}\right),\notag\\
    \iota_{\chi\wedge\bar{\chi}}\operatorname{hCS}(\bar{\mathcal{A}})=&-2\operatorname{tr}\left((\iota_{\bar\chi}\bar{\mathcal{A}})(\bar{\partial}+\bar{\mathcal{A}})(\iota_{\chi}\bar{\mathcal{A}})\right)+2\operatorname{tr}\left((\iota_{\bar\chi}\bar{\mathcal{A}})(\iota_{\chi}\bar{\mathcal{A}})\bar{\mathcal{A}}\right).\notag
\end{align}
Thus we have
\begin{align}
    \label{equ:1.19}
    &\iota_{\chi\wedge\bar{\chi}}\left(\Omega\wedge\operatorname{hCS}(\bar{\mathcal{A}})\right)
    =\langle\mathrm{d}\pi \pi\rangle\langle\pi\mu\rangle\langle\pi\hat{\mu}\rangle\Phi\operatorname{phCS}(\bar{A})
\end{align}
with
\begin{align}
    \label{equ:1.20}
    \bar{A}=\bar{e}^0\bar{\mathcal{A}}_0+ \left(\iota_{\omega}\bar{\mathcal{A}}-\frac{\langle\pi\hat{\mu}\rangle}{\langle\pi\mu\rangle}\iota_{\bar{\chi}}\bar{\mathcal{A}}\right)\mathrm{d}w+\left(\iota_{\bar{\omega}}\bar{\mathcal{A}}+\frac{\langle\pi\mu\rangle}{\langle\pi\hat{\mu}\rangle}\iota_{\chi}\bar{\mathcal{A}}\right)\mathrm{d}\bar{w},
\end{align}
and
\begin{align}
    \label{equ:1.21}
    \operatorname{phCS}(\bar{A})=\operatorname{tr}\left(\bar{A} \mathrm{d}^{\prime} \bar{A}+\frac{2}{3} \bar{A} \wedge \bar{A} \wedge \bar{A}\right),
\end{align}
where $\mathrm{d}^{\prime}=\bar{e}^{0} \bar{\partial}_{0}+\mathrm{d} w \partial_{w}+\mathrm{d} \bar{w} \partial_{\bar{w}}=\bar{\partial}_{\mathrm{CP}^{1}}+\mathrm{d}_{\mathbb{E}^{2}}$. In deriving \eqref{equ:1.19}, we have used the following identity 
\begin{align}
    \label{equ:1.22}
    \operatorname{hCS}(X+Y)=\operatorname{hCS}(X)+2 \operatorname{tr}(\bar{\mathcal{F}}(X) Y)-\bar{\partial} \operatorname{tr}(X Y)+2 \operatorname{tr}\left(X Y^{2}\right)+\operatorname{hCS}(Y). 
\end{align}
After the symmetry reduction, we end up with a 4DCS theory with action
\begin{align}
    \label{equ:1.23}
    S_{\omega}\left[\bar{A}\right]=\frac{1}{2 \pi i} \int_{\mathbb{E}^{2} \times \mathbb{C} \mathbb{P}^{1}} \omega \wedge \operatorname{phCS}\left(\bar{A}\right),
\end{align}
where $\omega=\langle\mathrm{d}\pi \pi\rangle\langle\pi\mu\rangle\langle\pi\hat{\mu}\rangle\Phi$ is a  meromorphic one-form. The 4D gauge connection $(\bar{A}_{w},\bar{A}_{\bar{w}})$ along $\mathbb{E}^{2}$ direction is related to the holomorphic gauge connection $\bar{\mathcal{A}}$   by
\begin{align}
    \label{add:1.17}
    \bar{A}_{w}=\iota_{\omega}\bar{\mathcal{A}}-\frac{\langle\pi\hat{\mu}\rangle}{\langle\pi\mu\rangle}\iota_{\bar{\chi}}\bar{\mathcal{A}}=-\frac{[\kappa\hat{\mathcal{A}}]}{\langle\pi\mu\rangle},\\
    \label{add:1.18}
    \bar{A}_{\bar{w}}=\iota_{\bar{\omega}}\bar{\mathcal{A}}+\frac{\langle\pi\mu\rangle}{\langle\pi\hat{\mu}\rangle}\iota_{\chi}\bar{\mathcal{A}}=-\frac{[\hat{\kappa}\hat{\mathcal{A}}]}{\langle\pi\hat{\mu}\rangle},
\end{align}
which we will refer to as the \textit{matching condition}. It relates the 4D gauge connection to the 6D gauge connection, if two theories are related simply by symmetry reduction. Then it is  natural to expect that the boundary conditions on the 4D gauge connection can be transferred to the ones on the 6D gauge connection.  

\subsection{Solving the 4D CS theory reduced from holomorphic Chern-Simons}
Taken $\Phi=(\langle\pi\alpha\rangle\langle\pi\beta\rangle)^{-2}$, the resulting one-form $\omega$ for the 4D CS theory is
\begin{align}
    \label{add:1.8}
    \omega=\frac{\langle\mathrm{d}\pi \pi\rangle\langle\pi\mu\rangle\langle\pi\hat{\mu}\rangle}{\langle\pi\alpha\rangle^2\langle\pi\beta\rangle^2}
\end{align}
with the normalization $\la \alpha\beta \ra=1$. 
Now we can follow the procedure proposed in \cite{CWY} to derive the 2D integrable field theories. At the two double poles the boundary term \eqref{add:1.7} can be set to zero by taking the Dirichlet boundary conditions
\begin{align}
    \label{add:1.9}
    \bar{A}_{w,\bar{w}}|_{\pi=\alpha}=0=\bar{A}_{w,\bar{w}}|_{\pi=\beta}.
\end{align}
The Lax connection $\mathcal{L}$ of the 2D integrable field theory is related to $\bar{A}$ by a gauge transformation
\begin{align}
    \label{equ:1.27}
    \bar{A}=\tilde{\sigma}^{-1}\mathrm{d}'\tilde{\sigma}+\tilde{\sigma}^{-1}\mathcal{L}\tilde{\sigma}. 
\end{align}
The choice of $\tilde{\sigma}$ has the gauge symmetry $\tilde{\sigma}\mapsto h\tilde{\sigma}g^{-1}$ where $h:\mathbb{E}^2\rightarrow G$ and $g:\mathbb{E}^2\times\mathbb{CP}^1\rightarrow G$. The freedom in $h$ corresponds to the gauge freedom in $\mathcal{L}$, and the freedom in $g$ corresponds to the gauge freedom in $\bar{A}$.  These gauge freedoms can be removed by fixing
\begin{align}
    \label{add:1.10}
    \tilde{\sigma}|_{\pi=\alpha}=\sigma,\qquad \tilde{\sigma}|_{\pi=\beta}=\operatorname{id}. 
\end{align}
Substituting \eqref{add:1.10} and \eqref{equ:1.27} into the boundary condition \eqref{add:1.9}, one finds
\begin{align}
    \label{add:1.11}
    \sigma^{-1}\mathrm{d}_{\mathbb{E}^2}\sigma+\sigma^{-1}\mathcal{L}|_{\pi=\alpha}\sigma=0, \\
    \label{add:1.12}
    \mathcal{L}|_{\pi=\beta}=0. 
\end{align}
At the positions of the zeros of the one-form $\omega$ one has to insert the defects which describe the pole structures of the 4D gauge fields such that the poles of the gauge field cancel the zeros of the one-form \cite{CWY}. The cancellation is necessary for the gauge field to have a non-degenerate propagator. Considering the condition \eqref{add:1.12}  the Lax connection has to be in the form
\begin{align}
    \label{add:1.13}
    \mathcal{L}_{w}=\frac{\langle\pi\beta\rangle}{\langle\pi\mu\rangle}U_{w},\qquad \mathcal{L}_{\bar{w}}=\frac{\langle\pi\beta\rangle}{\langle\pi\hat{\mu}\rangle}U_{\bar{w}} 
\end{align}
where $U_{w,\bar{w}}$ does not depend on $\pi^{A'}$. Putting \eqref{add:1.13} into \eqref{add:1.11},  one can solve
\begin{align}
    \label{add:1.14}
    \mathcal{L}_{w}=-\frac{\langle\pi\beta\rangle}{\langle\pi\mu\rangle}\langle\alpha\mu\rangle\partial_{w}\sigma\sigma^{-1}, \qquad \mathcal{L}_{\bar{w}}=-\frac{\langle\pi\beta\rangle}{\langle\pi\hat{\mu}\rangle}\langle\alpha\hat{\mu}\rangle\partial_{\bar{w}}\sigma\sigma^{-1}. 
\end{align}
Substituting \eqref{add:1.14} into \eqref{equ:1.27}  by using \eqref{equ:1.21} and \eqref{equ:1.23},  one ends up with the action of 2D WZW model 
\begin{align}
    \label{add:1.15}
    S_{\omega}\left[\bar{A}\right]=&\frac{1}{2 \pi i} \int_{\mathbb{E}^{2} \times \mathbb{C} \mathbb{P}^{1}} \omega \wedge \operatorname{phCS}\left(\bar{A}\right)\notag\\
    =& \int_{\mathbb{E}^{2}}  \operatorname{tr}(j_w j_{\bar{w}} )\mathrm{d}w\wedge\mathrm{d}\bar{w}
    +\frac{k}{3}\int_{\mathbb{E}^{2} \times [0,1]}\operatorname{tr}(\tilde{j}^3). 
\end{align}
where $\tilde{j}\equiv-\mathrm{d}'\tilde{\sigma}\tilde{\sigma}^{-1}$ and $k=\langle\alpha\hat{\mu}\rangle\langle\mu\beta\rangle+\langle\alpha\mu\rangle\langle\hat{\mu}\beta\rangle$.

\subsection{Solving along fibres for holomorphic Chern-Simons theory}
Next we derive the same 2D action \eqref{add:1.15} following the right route  in Fig. \ref{Figure2}. At the two double poles the 
boundary conditions of the 6D gauge connection can be read directly by using the boundary conditions of 4DCS \eqref{add:1.11} and \eqref{add:1.12} through the \textit{matching condition} \eqref{equ:1.20}
\begin{align}
    \label{add:1.16}
    [\kappa\hat{\mathcal{A}}|_{\pi=\alpha,\beta}]=0=[\hat{\kappa}\hat{\mathcal{A}}|_{\pi=\alpha,\beta}]. 
\end{align}
The $(3,0)$-form $\Omega$ here does not have a zero, thus the 6D gauge connection does not have any pole\footnote{In this case there is no need to insert any defects. But to be compatible with our later discussion, here we abuse the terminology by saying that we insert a trivial defect.} in $\mathbb{CP}^1$ such that $\bar{\mathcal{A}}_{A}=\pi^{A'}A_{AA'}$, where $A_{AA'}$ does not depend on the coordinates $\pi^{A'}$. By applying a formal gauge transformation the dynamical field
$\bar{\mathcal{A}}$ can be rewritten as
\begin{align}\label{4dLax} 
\bar{\mathcal{A}}=\hat{\sigma}^{-1}\bar{\partial}\hat{\sigma}+\hat{\sigma}^{-1}\bar{\mathcal{A}}'\hat{\sigma},
\end{align}
where $\bar{\mathcal{A}}'=\hat{e}^{A}\hat{\mathcal{A}}_{A}$ is the 4D analogue of the Lax connection. Similar to what we did in the 4DCS case, we can fix the gauge freedom by requiring that
\begin{align}
    \label{add:1.19}
    \hat{\sigma}|_{\pi=\alpha}=\sigma,\qquad \hat{\sigma}|_{\pi=\beta}=\operatorname{id},
\end{align}
and solve the boundary condition \eqref{add:1.16} 
\begin{align}
    \label{add:1.20}
    \bar{\mathcal{A}}'=\hat{e}^{A}\hat{\mathcal{A}}_{A}=-\langle\pi\beta\rangle\hat{e}^A\alpha^{A'}\partial_{AA'}\sigma\sigma^{-1}. 
\end{align}
Thus the resulting 4D action is given by
\begin{align}
    \label{add:1.21}
    S_{\Omega}[\bar{\mathcal{A}}]=&\frac{1}{2 \pi i} \int_{\mathbb{PT}} \Omega \wedge \mathrm{hCS}(\bar{\mathcal{A}})\notag\\
    =&\frac{1}{4\pi i}\sum_{z\in\mathfrak{p}} \lim _{\epsilon \rightarrow 0} \oint_{S_{z, \epsilon}^{1}}\left[\frac{\langle\mathrm{d} \pi \pi\rangle}{\langle\pi\alpha\rangle^2\langle\pi\beta\rangle^2} \int_{\mathbb{E}^{4}} \mathrm{d}^2 x^{A'B'}\pi_{A'}\pi_{B'}\wedge \operatorname{tr}(\bar{\mathcal{J}} \wedge \bar{\mathcal{A}}')\right]\notag\\
    &+\frac{1}{12\pi i}\int_{\mathbb{PT}}\frac{\langle\mathrm{d}\pi \pi\rangle\wedge\mathrm{d}^{2}x^{A'B'}\pi_{A'}\pi_{B'}}{\langle\pi\alpha\rangle^2\langle\pi\beta\rangle^2}\operatorname{tr}\left(\bar{\mathcal{J}^3}\right) 
\end{align}
where $\hat{\mathcal{J}}_A\equiv-\pi^{A'}\partial_{AA'}\hat{\sigma}\hat{\sigma}^{-1}$. 
Without losing any generality, we fix a normalization such that $\alpha^{A'}\beta^{B'}-\beta^{A'}\alpha^{B'}=\varepsilon^{A'B'}$ then \eqref{add:1.21} becomes 
\begin{align}
    \label{add:1.22}
    -\frac{1}{2}\int_{\mathbb{E}^4}\operatorname{vol}_4\varepsilon^{AB}\varepsilon^{A'B'}\operatorname{tr}(\partial_{AA'}\sigma\sigma^{-1}\partial_{BB'}\sigma\sigma^{-1})-\frac{1}{3}\int_{\mathbb{E}^4\times [0,1]}\mu_{\alpha,\beta}\wedge\operatorname{tr}(\tilde{J}^3) 
\end{align}
where $\mu_{\alpha,\beta}=\mathrm{d}^2x^{A'B'}\alpha_{A'}\beta_{B'}$ and $\tilde{J}=-\tilde{\mathrm{d}}\tilde{\sigma}\sigma^{-1}$ with $\tilde{\sigma}$ being a smooth homotopy from $\sigma$ to $\operatorname{id}$.  

\subsection{Symmetry reduction to 2D theory}

The next step is to apply the same symmetry reduction along $H$ as before to this 4D action \eqref{add:1.22} to get the action of the 2D integrable model. The actions of $H$ on $\mathbb{E}^4$ are generated by the vectors $X$ and $\bar{X}$ defined by the pullback
\begin{align}
    \label{1.60}
    \pi_*\chi=X, \qquad \pi_*\bar{\chi}=\bar{X}. 
\end{align}
Since $\chi$ and $\bar{\chi}$ do not depend on $\mathbb{CP}^1$, the expression for $X$ and $\bar{X}$ is  the same as before, i.e.
\begin{align}
    \label{1.61}
    X=\hat{\kappa}^{A} \hat{\mu}^{A^{\prime}} \partial_{A A^{\prime}}=\partial_{z}, \qquad \bar{X}=\kappa^{A} \mu^{A^{\prime}} \partial_{A A^{\prime}}=\partial_{\bar{z}}. 
\end{align}
Now we impose the constraint that the dynamical field $\sigma$ for this 4D effective model has the same symmetry
\begin{align}
    \label{1.62}
    \mathcal{L}_{X}\sigma=\mathcal{L}_{\bar{X}}\sigma=0.  
\end{align}
Introducing the following vector fields
\begin{align}
    \partial_{w}=\kappa^A\hat{\mu}^{A'}\partial_{AA'},\quad \partial_{\bar{w}}=-\hat{\kappa}^A\mu^{A'}\partial_{AA'},\quad \partial_{z}=\hat{\kappa}^A\hat{\mu}^{A'}\partial_{AA'},\quad \partial_{\bar{z}}=\kappa^A\mu^{A'}\partial_{AA'}, \label{add:1.23}
\end{align}
the symmetry reduction of \eqref{add:1.22} can be easily performed as in the 6D case. The resulting action is
\begin{align} 
    \label{add:1.24}
    \int_{\mathbb{E}^{2}}  \operatorname{tr}(j_w j_{\bar{w}} )\mathrm{d}w\wedge\mathrm{d}\bar{w}
    +\frac{k}{3}\int_{\mathbb{E}^{2} \times [0,1]}\operatorname{tr}(\tilde{j}^3),
\end{align}
which is exactly the same as what we got by directly solving the reduced 4DCS theory. Moreover, we find that the Lax connections derived from the two different routes indeed match with each other
\begin{align}
    \label{add:1.25}
    \mathcal{L}_{w}=-\frac{[\kappa\hat{\mathcal{A}}']}{\langle\pi\mu\rangle}=\frac{\langle\pi\beta\rangle}{\langle\pi\mu\rangle}\kappa^A\alpha^{A'}\partial_{AA'}\sigma\sigma^{-1}=-\frac{\langle\pi\beta\rangle}{\langle\pi\mu\rangle}\langle\alpha\mu\rangle\partial_w \sigma\sigma^{-1},\\
    \label{add:1.26}
    \mathcal{L}_{\bar{w}}=-\frac{[\hat{\kappa}\hat{\mathcal{A}}]}{\langle\pi\hat{\mu}\rangle}=\frac{\langle\pi\beta\rangle}{\langle\pi\hat{\mu}\rangle}\hat{\kappa}^A\alpha^{A'}\partial_{AA'}\sigma\sigma^{-1}=-\frac{\langle\pi\beta\rangle}{\langle\pi\hat{\mu}\rangle}\langle\alpha\hat{\mu}\rangle\partial_{\bar{w}}\sigma\sigma^{-1}. 
\end{align}
This concludes the commutativity of the diagram in Fig. \ref{Figure2}.

It is very instructive to derive the Lax connection of the 2D model directly from ASDYM
\bea\label{ASDYM}
\epsilon^{AB}[\nabla_{AA'},\nabla_{BB'}]=0.
\eea 
Rewriting the 4D gauge Lax connection  \eqref{add:1.20} as $A_{AA'}=\beta_{A'}A_A,~ A_A=\alpha^{B'}J_{AB'}$ then \eqref{ASDYM} becomes
\bea \label{ASDYM1}
\mathcal{E}_{A'B'}=\epsilon^{AB}\(\beta_{B'}\p_{AA'}A_B-\beta_{A'}\p_{BB'}A_A+\beta_{A'}\beta_{B'}[A_A,A_B]\).
\eea 
Since $\la \alpha\beta\ra=1$ we can choose $(\alpha^{A'},\beta^{A'})$ as the dyads therefore \eqref{ASDYM1} is equivalent to three equations obtained by contracting it with $\alpha^{A'}\alpha^{B'}$, $\beta^{A'}\beta^{B'}$ and $\beta^{A'}\alpha^{B'}$ separately 
\bea 
&&\beta^{A'}\beta^{B'}\mathcal{E}_{A'B'}=0, \label{Con1}\\
&&\alpha^{A'}\alpha^{B'}\mathcal{E}_{A'B'}=\epsilon^{AB}\(\alpha^{A'}\p_{AA'}A_B-\alpha^{B'}\p_{BB'}A_A+[A_A,A_B]\)=0,\label{Con2}\\
&&\beta^{A'}\alpha^{B'}\mathcal{E}_{A'B'}=\epsilon^{AB}\beta^{A'}\p_{AA'}A_B=0.\label{Con3}
\eea 
Performing symmetry reduction on \eqref{Con3} gives the equation of motion \eqref{add:1.24} of 2D theory 
\bea 
\epsilon^{AB}\beta^{A'}\p_{AA'}A_B&=&\la\beta\mu\ra\la \alpha\hat{\mu}\ra\p_w J_{\bar{w}}-\la\beta\hat{\mu}\ra\la \alpha{\mu}\ra\p_{\bar{w}} J_{{w}}\nn
&=&(1-k)\p_w J_{\bar{w}}+(1+k)\p_{\bar{w}}J_{w}=0,\label{Eom}
\eea 
where in the last line we have used the following identities
\bea 
&&\la\alpha\beta\ra=\la\alpha \hat{\mu}\ra\la\mu \beta\ra-\la\alpha {\mu}\ra\la\hat{\mu} \beta\ra=1,\\
&&\la\alpha \hat{\mu}\ra\la\mu \beta\ra+\la\alpha {\mu}\ra\la\hat{\mu} \beta\ra=-k.
\eea
Similarly the symmetry reduction of \eqref{Con2} gives the flatness condition 
\bea 
\p_w J_{\bar{w}}-\p_{\bar{w}} J_{{w}}+[J_{w},J_{\bar{w}}]=0.
\eea 
Therefore the 2D Lax connection can be obtained by contracting $(\alpha^{A'}+\lambda \beta^{A'})(\alpha^{B'}+\lambda \beta^{B'})$ with \eqref{ASDYM} and doing the symmetry reduction
\bea 
(\alpha^{A'}+\lambda \beta^{A'})(\alpha^{B'}+\lambda \beta^{B'})\mathcal{E_{A'B'}}\sim [\p_w+\frac{\la\alpha \mu\ra}{\la\alpha\mu\ra+\lambda \la\beta\mu\ra}J_{w},\p_{\bar{w}}+\frac{\la\alpha \hat{\mu}\ra}{\la\alpha\hat{\mu}\ra+\lambda \la\beta\hat{\mu}\ra}J_{\bar{w}}],\notag
\eea 
where $\lambda\in \mathbb{C}$ is the spectral parameter.  Comparing with \eqref{add:1.25} and  \eqref{add:1.26}, one can find
\bea 
\alpha^{A'}+\lambda \beta^{A'}=-\frac{\pi^{A'}}{\la \pi \beta\ra}~~\Rightarrow ~~\lambda=\frac{\la \pi \alpha\ra}{\la \pi \beta\ra}\notag
\eea 
to get the agreement.

\section{$\lambda$-deformed principal chiral model from holomorphic Chern-Simons}

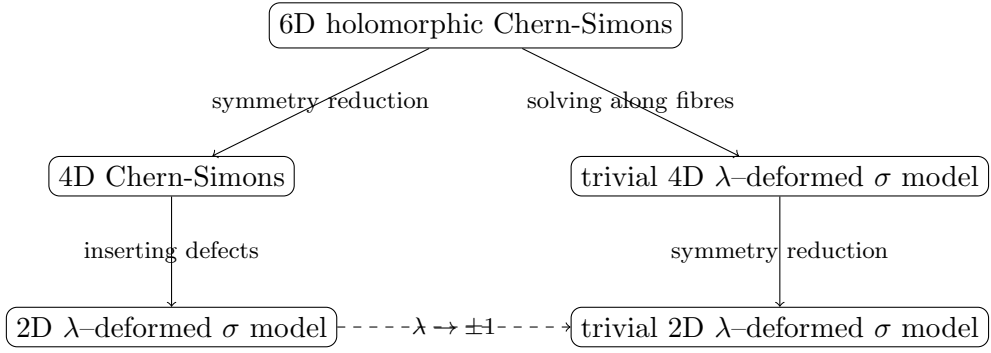
\begin{figure}[h]
\label{Figure3}
\begin{center}
\begin{tikzpicture}[scale=1]
\path (-4,0) node[rectangle,
           rounded corners,draw](2D1)  {2D $\lambda$--deformed $\sigma$ model}
           (4,0) node[rectangle,
           rounded corners,draw](2D2)  {trivial 2D $\lambda$--deformed $\sigma$ model}
(0,4) node[rectangle,
           rounded corners,draw] (6D) {6D holomorphic Chern-Simons}
(-4,2) node[rectangle,
           rounded corners,draw] (4DCS) {4D Chern-Simons}
(4,2) node[rectangle,
           rounded corners,draw] (4DI) {trivial 4D $\lambda$--deformed  $\sigma$ model};
\draw[->] (6D) -- (4DCS)
 node[pos=0.5]{{\footnotesize symmetry reduction}};
\draw[->] (6D) -- (4DI)
node[pos=0.5]{{\footnotesize solving along fibres }};
\draw[->] (4DCS) -- (2D1)
node[pos=0.5]{{\footnotesize inserting defects}};
\draw[->] (4DI) -- (2D2)
node[pos=0.5]{{\footnotesize symmetry reduction}};
\draw[dashed,->](2D1)--(2D2)
node[pos=0.5]{{\footnotesize $\lambda\rightarrow \pm 1$}};
\end{tikzpicture}
\end{center}
\caption{A summary of the construction of $\lambda$ deformation. }
\end{figure}

In this section, we will show that the diagram in Fig. \ref{Figure2} can not be simply generalized to the $\lambda$-deformed theory. Instead, we find the following construction summarized in the diagram in Fig. \ref{Figure3}.

The $\lambda$-deformed principal chiral model is derived from 4DCS with a 1-form 
\begin{align}
    \label{equ:1.24}
    \omega=\frac{\langle\mathrm{d}\pi \pi\rangle\langle\pi\mu\rangle\langle\pi\hat{\mu}\rangle}{\langle\pi\alpha_+\rangle\langle\pi\alpha_-\rangle\langle\pi\beta\rangle^2}. 
\end{align}
The relation \eqref{equ:1.19} between $\omega$ and $\Omega$ implies that the $(3,0)$-form $\Omega$ should be  
\begin{align}
    \label{equ:1.4}
    \Omega=\frac{\langle\mathrm{d}\pi\pi\rangle\wedge\mathrm{d}^2x^{A'B'}\pi_{A'}\pi_{B'}}{2\langle\pi\alpha_+\rangle\langle\pi\alpha_-\rangle\langle\pi\beta\rangle^2}. 
\end{align}
As we discuss in the last section, varying $S_\Omega$ gives rise to the following boundary terms
\begin{align}
\label{equ:1.7}
    \delta S_{\Omega}|_{bdy}
    =&-\frac{1}{2 \pi i} \sum_{z\in\mathfrak{p}} \lim _{\epsilon \rightarrow 0} \oint_{S_{z, \epsilon}^{1}}\left[\frac{\langle\mathrm{d} \pi \pi\rangle}{\langle\pi\alpha_+\rangle\langle\pi\alpha_-\rangle\langle\pi\beta\rangle^2} \int_{\mathbb{E}^{4}} \mathrm{vol}_4 \varepsilon^{AB} \operatorname{tr}(\delta \bar{\mathcal{A}_A} \bar{\mathcal{A}}_B)\right],
\end{align}
which should vanish by imposing appropriate boundary conditions. We will follow the left route first, and then demonstrate that the matching conditions \eqref{add:1.17} and \eqref{add:1.18} will not lead to a well defined boundary conditions for 6DhCS such that the diagram in Fig.\ref{Figure3} is not commutative.

\subsection{Solving the 4D CS theory reduced from holomorphic Chern-Simons}
In \eqref{equ:1.4} we assume that the residues at the two simple poles are opposite, i.e. 
\begin{align}
    \label{equ:a.1}
    \frac{\langle\alpha_+\mu\rangle\langle\alpha_+\hat{\mu}\rangle}{\langle\alpha_+\beta\rangle^2}=\frac{\langle\alpha_-\mu\rangle\langle\alpha_-\hat{\mu}\rangle}{\langle\alpha_-\beta\rangle^2},
\end{align}
then the 1-form coincides with the choice of 1-form made in \cite{Delduc:2019whp} to derive the $\lambda$ deformation. At the boundary associated with the double pole, we still impose the Dirichelett boundary condition. The boundary conditions at the two simple poles are more tricky. The vanishing condition of the boundary terms at the two simple poles is given by
\bea 
\text{Res}_{\alpha_+}\omega\(\operatorname{tr}(\bar{A},\delta \bar{A})|_{\alpha_+}-\operatorname{tr}(\bar{A},\delta \bar{A})|_{\alpha_-}\)=0,
\eea 
where we have used $\text{Res}_{\alpha_+}\omega=-\text{Res}_{\alpha_-}\omega$.
This condition can be solved in general by requiring that the pair $(\bar{A}|_{\alpha_-},\bar{A}|_{\alpha_-})$ takes values in a Lagrangian subalgebra of this two-copy algebra $(\mathfrak{g},\mathfrak{g})\equiv \mathfrak{d}$, where $\mathfrak{g}$ is algebra of the gauge group $G$. To derive the $\lambda$-deformation, the Lagrangian subalgebra is $\mathfrak{g}^\delta=(x,x),x\in\mathfrak{g}$. If $\mathfrak{g}$ is a semi-simple complex algebra, then the complementary part $\mathfrak{g}_R$ of $\mathfrak{g}^\delta$ in $\mathfrak{d}$ is also a Lagrangian subalgebra and the triplet $(\mathfrak{d},\mathfrak{g}^\delta,\mathfrak{g}_R)$ forms a Manin triple. Therefore in this case, the boundary conditions should be
\begin{align}
    \label{equ:1.25}
    \left.\bar{A}_{w,\bar{w}}\right|_{\pi=\alpha_+}=&\left.\bar{A}_{w,\bar{w}}\right|_{\pi=\alpha_-},\\
    \label{equ:1.26}
    \left.\bar{A}_{w,\bar{w}}\right|_{\pi=\beta}=&0. 
\end{align}

The Lax connection $\mathcal{L}$ for 2D integrable field theory is related to $\bar{A}$ by the gauge transformation \eqref{4dLax}. As shown in \cite{Delduc:2019whp}, the gauge parameter $\tilde{\sigma}$  satisfies the so-called \textit{archipelago conditions}, which we omit the details here. The important fact about the  \textit{archipelago conditions}  is that $\tilde{\sigma}$ can be chosen to be the identity almost everywhere expect for  the neighbors around each poles, where they should be 
\bea\label{GaugeFreedom}
 \tilde{\sigma}|_{\alpha_+}\equiv\sigma,\qquad \tilde{\sigma}|_{\alpha_-}=\sigma_-,\qquad \tilde{\sigma}|_{\beta}=\sigma_\beta. 
\eea 
Next using the gauge symmetry $\tilde{\sigma}\mapsto h\tilde{\sigma},h:\mathbb{E}^2\rightarrow G$ one can set $\sigma_\beta=\on{id}$, and using the residue gauge symmetry at the two simple poles $\tilde{\sigma}\mapsto \tilde{\sigma}g^{-1}, g \in (G,G)$ one can set $\sigma_{-}=\on{id}$.
After fixing all the gauge freedoms, the boundary conditions \eqref{equ:1.25} and \eqref{equ:1.26} are equivalently rewritten as
\begin{align}
    \label{equ:1.29}
    (\sigma^{-1}\mathrm{d}_{\mathbb{E}^2}\sigma+\sigma^{-1}\mathcal{L}|_{\pi=\alpha_+}\sigma)=&\mathcal{L}|_{\pi=\alpha_-},\\
    \label{equ:1.30}
    \mathcal{L}|_{\pi=\beta}=&0. 
\end{align}
Since the 1-form has two zeros, the Lax connection has to be in the form \eqref{add:1.13}.
The boundary condition \eqref{equ:1.29} becomes 
\begin{align}
    \label{equ:1.33}
    \left(\frac{\langle\alpha_+\beta\rangle}{\langle\alpha_+\hat{\mu}\rangle}-\frac{\langle\alpha_-\beta\rangle}{\langle\alpha_-\hat{\mu}\rangle}\operatorname{Ad}_{\sigma}\right)U_{\bar{w}}=&-\partial_{\bar{w}}\sigma \sigma^{-1},\\
    \label{equ:1.34}
    \left(\frac{\langle\alpha_+\beta\rangle}{\langle\alpha_+\mu\rangle}-\frac{\langle\alpha_-\beta\rangle}{\langle\alpha_-\mu\rangle}\operatorname{Ad}_{\sigma}\right)U_{w}=&-\partial_{w}\sigma \sigma^{-1},
\end{align}
which can be simply solved and the resulting  2D Lax connection is just
\begin{align}
    \label{equ:1.37}
    \mathcal{L}_{\bar{w}}=\frac{\langle\pi\beta\rangle}{\langle\pi\hat{\mu}\rangle}\frac{\langle\alpha_+\hat{\mu}\rangle}{\langle\alpha_+\beta\rangle}\frac{1}{\left(1-\frac{\langle\alpha_-\beta\rangle\langle\alpha_+\hat{\mu}\rangle}{\langle\alpha_+\beta\rangle\langle\alpha_-\hat{\mu}\rangle}\operatorname{Ad}_{\sigma}\right)}j_{\bar{w}},\\
    \label{equ:1.38}
    \mathcal{L}_{w}=\frac{\langle\pi\beta\rangle}{\langle\pi\mu\rangle}\frac{\langle\alpha_+\mu\rangle}{\langle\alpha_+\beta\rangle}\frac{1}{\left(1-\frac{\langle\alpha_-\beta\rangle\langle\alpha_+\mu\rangle}{\langle\alpha_+\beta\rangle\langle\alpha_-\mu\rangle}\operatorname{Ad}_{\sigma}\right)}j_{w}. 
\end{align}
Substituting \eqref{equ:1.37} into \eqref{equ:1.27}  by using \eqref{equ:1.21} and \eqref{equ:1.23},  we end up with the action of $\lambda$--deformed principal chiral model 
\begin{align}
    \label{equ:1.39}
    S_{\omega}\left[\bar{A}\right]=&\frac{1}{2 \pi i} \int_{\mathbb{E}^{2} \times \mathbb{C} \mathbb{P}^{1}} \omega \wedge \operatorname{phCS}\left(\bar{A}\right)\notag\\
    =&\frac{1}{2 \pi i}\sum_{z\in\mathfrak{p}} \lim _{\epsilon \rightarrow 0} \oint_{S_{z, \epsilon}^{1}}\left[\frac{\langle\mathrm{d} \pi \pi\rangle\langle\pi\mu\rangle\langle\pi\hat{\mu}\rangle}{\langle\pi\alpha_+\rangle\langle\pi\alpha_-\rangle\langle\pi\beta\rangle^2} \int_{\mathbb{E}^{2}}  \operatorname{tr}(j \wedge \mathcal{L})\right]\notag\\
    &+\frac{1}{6\pi i}\int_{\mathbb{E}^{2} \times \mathbb{C} \mathbb{P}^{1}}\frac{\langle\mathrm{d} \pi \pi\rangle\langle\pi\mu\rangle\langle\pi\hat{\mu}\rangle}{\langle\pi\alpha_+\rangle\langle\pi\alpha_-\rangle\langle\pi\beta\rangle^2}\operatorname{tr}(\tilde{j}^3),
\end{align}
where the first term of \eqref{equ:1.39} explicitly reads
\begin{align}
    \label{equ:1.40}
    &\frac{1}{2 \pi i}\sum_{z\in\mathfrak{p}} \lim _{\epsilon \rightarrow 0} \oint_{S_{z, \epsilon}^{1}}\left[\frac{\langle\mathrm{d} \pi \pi\rangle\langle\pi\mu\rangle\langle\pi\hat{\mu}\rangle}{\langle\pi\alpha_+\rangle\langle\pi\alpha_-\rangle\langle\pi\beta\rangle^2} \int_{\mathbb{E}^{2}}  \operatorname{tr}(j \wedge \mathcal{L})\right]\notag\\
    =&\frac{\langle\alpha_+\mu\rangle\langle\alpha_+\hat{\mu}\rangle}{\langle\alpha_+\alpha_-\rangle\langle\alpha_+\beta\rangle^2}\int_{\mathbb{E}^{2}}\left(\operatorname{tr}(\sigma^{-1}\partial_w\sigma\sigma^{-1}\partial_{\bar{w}}\sigma)+2\operatorname{tr}(\frac{1}{\lambda-\operatorname{Ad}_{\sigma^{-1}}}\sigma^{-1}\partial_{w}\sigma\partial_{\bar{w}}\sigma\sigma^{-1})\right)\mathrm{d}w\wedge\mathrm{d}\bar{w},
\end{align}
and the second term is just the topological term. The deformation parameter is defined to be  
\be
\lambda\equiv\frac{\langle\alpha_-\beta\rangle\langle\alpha_+\mu\rangle}{\langle\alpha_+\beta\rangle\langle\alpha_-\mu\rangle}=\frac{\langle\alpha_+\beta\rangle\langle\alpha_-\hat{\mu}\rangle}{\langle\alpha_-\beta\rangle\langle\alpha_+\hat{\mu}\rangle}.
\ee
% thanks to \eqref{equ:a.1}. 

 \subsection{Boundary conditions for holomorphic Chern-Simons theory}
If one assumes that the commutativity in Fig. \ref{Figure1} still holds, then the boundary conditions \eqref{equ:1.25}  for 4DCS theory can be lifted to the boundary conditions for 6DhCS via the \textit{matching conditions} \eqref{add:1.17} and \eqref{add:1.18} which implies that the boundary condition on the 6D gauge connection $\hat{\mathcal{A}}$ should satisfy
\begin{align}
    \label{equ:1.45}
    \langle\alpha_-\mu\rangle[\kappa \hat{\mathcal{A}}]|_{\pi=\alpha_+}=\langle\alpha_+\mu\rangle[\kappa \hat{\mathcal{A}}]|_{\pi=\alpha_-},\\
    \label{equ:1.46}
    \langle\alpha_-\hat{\mu}\rangle[\hat{\kappa}\hat{\mathcal{A}}]|_{\pi=\alpha_+}=\langle\alpha_+\hat{\mu},\rangle[\hat{\kappa}\hat{\mathcal{A}}]|_{\pi=\alpha_-}. 
\end{align}
This boundary condition itself is well defined in the sense that it makes the boundary terms \eqref{equ:1.7} vanish. But it is problematic to construct a 4D theory via solving along fibres. Because the appearance of the extra factors, the pair $(\hat{\mathcal{A}}|_{\alpha_+},\hat{\mathcal{A}}|_{\alpha_-})$ is not in the Lagrangian subalgebra so that the residue gauge freedoms in \eqref{GaugeFreedom} can not be removed completely\footnote{Probably the gauge freedoms can be removed by introducing auxiliary edge modes as described in \cite{Benini:2020skc}.}. Therefore the resulting 4D theory and 2D theory would be some field theories with gauge symmetries, rather than  the $\lambda$-deformed theories that we expect. To get a boundary condition without such problem, we have to set $\frac{\langle\alpha_-\mu\rangle}{\langle\alpha_-\hat{\mu}\rangle}=\frac{\langle\alpha_+\mu\rangle}{\langle\alpha_+\hat{\mu}\rangle}$ which means
\begin{align}
    \langle\alpha_-\hat{\mu}\rangle\langle\mu\alpha_+\rangle-\langle\alpha_-\mu\rangle\langle\hat{\mu}\alpha_+\rangle=\langle\alpha_-\alpha_+\rangle=0. 
\end{align}
However this implies that $\alpha_-$ and $\alpha_+$ collide to a double pole such that  \eqref{equ:1.24} reduces to \eqref{add:1.8}, which leads to an un-deformed theory.
Given the failure of the matching condition one may wonder how about directly imposing $(\hat{\mathcal{A}}|_{\alpha_+},\hat{\mathcal{A}}|_{\alpha_-})\in \mathfrak{g}^\delta$. Next we show this  will indeed lead to $\lambda$ deformation but only with $\lambda=\pm 1$.

\subsection{Solving along fibres for holomorphic Chern-Simons theory}
Recall that the $(3,0)$-form $\Omega$ is 
\begin{align}
    \label{Omega}
    \Omega=\frac{\langle\mathrm{d}\pi\pi\rangle\wedge\mathrm{d}^2x^{A'B'}\pi_{A'}\pi_{B'}}{2\langle\pi\alpha_+\rangle\langle\pi\alpha_-\rangle\langle\pi\beta\rangle^2}.
\end{align}
To impose $(\hat{\mathcal{A}}|_{\alpha_+},\hat{\mathcal{A}}|_{\alpha_-})\in \mathfrak{g}^\delta$,  we need to set
\bea \label{Constr}
\la \alpha_+ \beta\ra^2=\la \alpha_-\beta\ra^2,
\eea 
such that the residues at the two simple poles are opposite. Therefore the boundary conditions are 
\bea 
\label{Bdy1}
\hat{\mathcal{A}}|_\beta=0,\quad \hat{\mathcal{A}}|_{\alpha_+}=\hat{\mathcal{A}}|_{\alpha_-}.
\eea 
The 4D Lax connection is given by \eqref{4dLax} and can be written as
\bea \label{GaugeField}
\hat{\mathcal{A}}'_A=\pi^{A'}A_{AA'}.
\eea 
The gauge parameter $\hat{\sigma}$ now can be fixed to be
\bea\label{archipelago}
\hat{\sigma}|_{\pi=\alpha_-}=\hat{\sigma}|_{\pi=\beta}=\operatorname{id},\quad \hat{\sigma}|_{\pi=\alpha_+}\equiv\sigma. 
\eea
Substituting it into the first boundary condition in \eqref{Bdy1}, we find
\bea 
\beta^{A'}A_{AA'}=0,\quad \Rightarrow \quad A_{AA'}\sim  \beta_{A'},\notag
\eea
which implies
\bea 
\hat{\mcal{A}}'_A=\la \pi \beta\ra A_A.
\eea 
The second boundary condition in \eqref{Bdy1} then gives the relation
\bea 
\la \alpha_-\beta\rangle A_A=\sigma^{-1} \alpha_+^{A'}\p_{AA'}\sigma+\la \alpha_+\beta\ra \sigma^{-1}A_A\sigma,\notag
\eea 
from which we read
\bea 
A_A=-\frac{\alpha_+^{A'}\p_{AA'}\sigma \sigma^{-1}}{\la\alpha_+\beta\ra \(1-\frac{\la \alpha_-\beta\ra}{\la\alpha_+\beta\ra}\text{Ad}_\sigma\)}.
\eea 
Substituting them into the 6D action and defining $\bar{\mathcal{J}}\equiv -\bar{\p}\hat{\sigma}\hat{\sigma}^{-1}$, we have
\begin{align}
    \label{6Daction}
    S_{\Omega}[\bar{\mathcal{A}}]=&\frac{1}{2 \pi i} \int_{\mathbb{PT}} \Omega \wedge \mathrm{hCS}(\bar{\mathcal{A}})\notag\\
    =&\frac{1}{4\pi i}\sum_{z\in\mathfrak{p}} \lim _{\epsilon \rightarrow 0} \oint_{S_{z, \epsilon}^{1}}\left[\frac{\langle\mathrm{d} \pi \pi\rangle}{\langle\pi\alpha_+\rangle\langle\pi\alpha_-\rangle\langle\pi\beta\rangle^2} \int_{\mathbb{E}^{4}} \mathrm{d}^2 x^{A'B'}\pi_{A'}\pi_{B'}\wedge \operatorname{tr}(\bar{\mathcal{J}} \wedge \bar{\mathcal{A}}')\right]\notag\\
    &+\frac{1}{12\pi i}\int_{\mathbb{PT}}\frac{\langle\mathrm{d}\pi \pi\rangle\wedge\mathrm{d}^{2}x^{A'B'}\pi_{A'}\pi_{B'}}{\langle\pi\alpha_+\rangle\langle\pi\alpha_-\rangle\langle\pi\beta\rangle^2}\operatorname{tr}\left(\bar{\mathcal{J}}^3\right).\end{align}
The kinetic term in \eqref{6Daction} is
\begin{align}
    \label{Kterm}
    &\frac{1}{4\pi i}\sum_{z\in\mathfrak{p}} \lim _{\epsilon \rightarrow 0} \oint_{S_{z, \epsilon}^{1}}\left[\frac{\langle\mathrm{d} \pi \pi\rangle}{\langle\pi\alpha_+\rangle\langle\pi\alpha_-\rangle\langle\pi\beta\rangle^2} \int_{\mathbb{E}^{4}} \mathrm{d}^2 x^{A'B'}\pi_{A'}\pi_{B'}\wedge \operatorname{tr}(\bar{\mathcal{J}} \wedge \bar{\mathcal{A}}')\right]\notag\\
    =&-\frac{\alpha_+^{A'}\alpha_+^{B'}}{2\langle\alpha_+\alpha_-\rangle\langle\alpha_+\beta\rangle^2} \varepsilon^{AB} \int_{\mathbb{E}^{4}}\operatorname{vol}_4  \operatorname{tr}\(\bar{\mathcal{J}}_{AA'}(\frac{1}{1-\lambda \operatorname{Ad}_\sigma}-\frac{1}{1-\lambda^{-1} \operatorname{Ad}_\sigma^{-1}}) \bar{\mathcal{J}}_{BB'}\),
\end{align} 
where we used the identity \eqref{equ:1.6}
 and $\lambda=\frac{\la \alpha_-\beta\ra}{\la\alpha_+\beta\ra}$, with 
 \be
 \lambda^2=1. \label{lambda1}\ee 
The topological term in the action \eqref{6Daction} becomes
\begin{align}
    \label{WZ}
    \frac{1}{12\pi i}\int_{\mathbb{PT}}\frac{\langle\mathrm{d}\pi \pi\rangle\wedge\mathrm{d}^{2}x^{A'B'}\pi_{A'}\pi_{B'}}{\langle\pi\alpha_+\rangle\langle\pi\alpha_-\rangle\langle\pi\beta\rangle^2}\operatorname{tr}\left(\bar{\mathcal{J}}^3\right)=\frac{1}{6}\frac{1}{\langle\alpha_+\alpha_-\rangle\langle\alpha_+\beta\rangle^2}\int_{\mathbb{E}^4\times[0,1]}\mu_{\alpha_+,\alpha_+}\wedge\operatorname{tr}\left(\tilde{\mathcal{J}}_{}^3\right)
\end{align}
where $\mu_{\alpha_+,\alpha_+}\equiv\mathrm{d}^2x^{A'B'}\alpha_+^{A'}\alpha_+^{B'}$. Combining \eqref{Kterm} with \eqref{WZ}, we get the 4D $\lambda$-deformed WZW model. Now imposing the symmetry  $H$ to the 4D field $\sigma$  as we did in the last section, we can get the 2D action with the kinetic term
\begin{align}
    \label{equ:1.68}
    \frac{\langle\alpha_+\mu\rangle\langle\alpha_+\hat{\mu}\rangle}{\langle\alpha_+\alpha_-\rangle\langle\alpha_+\beta\rangle^2}\int_{\mathbb{E}^{2}}\left(\operatorname{tr}(\partial_{w}\sigma\sigma^{-1}\partial_{\bar{w}}\sigma\sigma^{-1})+2\operatorname{tr}\left(\frac{\operatorname{Ad}_{\sigma^{-1}}}{\lambda-\operatorname{Ad}_{\sigma^{-1}}}\partial_{w}\sigma\sigma^{-1}\partial_{\bar{w}}\sigma\sigma^{-1}\right)\right)\mathrm{d}w\wedge\mathrm{d}\bar{w},
\end{align}
and the 3D topological term 
%By similar analysis we have
\begin{align}
    \label{equ:1.69}
    &\iota_{\chi\wedge\bar{\chi}}\left(\mathrm{d}^{2} x^{A^{\prime} B^{\prime}} \alpha_{+A^{\prime}} \alpha_{+B^{\prime}} \wedge \bar{\mathcal{J}}_{\alpha_+}^3\right)
    =2\langle\alpha_+\mu\rangle\langle\alpha_+\hat{\mu}\rangle \bar{\mathcal{J}}_{}^3.
\end{align}
However  the parameter $\lambda$ is not a continuous parameter anymore, instead $\lambda=\pm 1$ which makes the deformation less interesting. What is worse is that the operator $1-\operatorname{Ad}_\sigma$ may not be invertible considering $\operatorname{Ad}_\sigma \operatorname{Ad}_\sigma^T=1 $. 

The failure of the construction could be understood from different points of view. One way to see it is by counting the number of free parameters. In the $(3,0)$-form $\Omega$ there are only  two parameters. The boundary condition fixes one of the free parameters so there is only one parameter left over in the 4D action which is the overall coupling factor. Another way to understand the failure is from the fact that because there is no zero in the $(3,0)$-form $\Omega$ there is actually no non-trivial defect  inserted! Besides, if we take the $\lambda$-model as a special case of $\mathcal{E}$-model then there is another hint of the failure. In \cite{Lacroix:2020flf}, a large class of integrable $\mathcal{E}$-models was constructed from the 4DCS theory. The construction relies on the fact that the difference between the numbers of the zeros and the numbers of poles is two. However in the 6DhCS the difference is actually 4. So the strategy introduced in \cite{Lacroix:2020flf} seems not applicable directly. 

\section{Other possible deformations}
In this section, we study some other boundary conditions which have been used in 4DCS to construct 2D integrable field theories, including the $
\eta$--deformation and the generalized $\lambda$--deformation \cite{Gen} and the deformation of a coupled theory. In the first two cases, because there is no zero in the $(3,0)$-form $\Omega$ so that only trivial defects are inserted. In none of these two cases we get our desired deformation since the deformation parameters are strictly restricted, compared to those studied in the literature.

\subsection{$\eta$-deformation}\label{4.1}

When we discussed $\lambda$--deformation in the last section, we have used one of the Lagrangian subalgebra $\mathfrak{g}^\delta$ in the Manin triple $(\mathfrak{d},\mathfrak{g}^\delta,\mathfrak{g}_R)$. In the context of 4DCS, the other Lagrangian subalgebra $\mathfrak{g}_R$ will lead to $\eta$--deformation (the Yang-Baxter deformation). This choice of Lagrangian subalgebra requires 
the following boundary condition at the simple poles,
\bea 
\label{Bdy2}
\quad (R-i)\hat{\mathcal{A}}|_{\alpha_+}=(R+i)\hat{\mathcal{A}}|_{\alpha_-},
\eea 
where the skew symmetric operator $R$ satisfies the (non-split type) modified classical Yang-Baxter equation:
\bea 
[Rx,Ry]-R\([Rx,y]+[x,Ry]\)=[x,y],\quad \(x,y\in\mathfrak{g},\quad R\in \mbox{End }\mathfrak{g} \).
\eea 
In this case the gauge freedoms can be removed by fixing $\hat{\sigma}$ to be
\bea\label{fbdy2}
\hat{\sigma}|_{\pi=\alpha_\pm}=\sigma,\quad \hat{\sigma}|_{\pi=\beta}\equiv \operatorname{id}.
\eea 
Thus, after assuming \eqref{GaugeField}, we get the following equations
\bea 
&&\hat{\mathcal{A}}'_A=\la \pi\beta\ra A_A,\nn
&&(R_\sigma-i)(-J_A^{\alpha_+}+\la \alpha_+\beta\ra A_A)=(R_\sigma+i)(-J_A^{\alpha_-}+\la\alpha_-\beta\ra A_A),
\eea 
where $R_\sigma=\sigma R\sigma^{-1}$. The solution of these equations is
\bea 
\label{eta4d}
&&A_A=\frac{-R_\sigma (J_A^{\alpha_+}-J_A^{\alpha_-})+i (J_A^{\alpha_+}+J_A^{\alpha_-})}{-R_\sigma \la\gamma_-\beta\ra+i \la\gamma_+\beta\ra},\quad \gamma_\pm\equiv \alpha_+\pm \alpha_-.
\eea
To avoid the singularity of $R^{-1}$, we set $\la\gamma_-\beta\ra=0$ and leave $\la\gamma_+\beta\ra$ to be arbitrary. Substituting \eqref{eta4d} into \eqref{6Daction}, we find that the action of the 4D theory is
\bea 
S_{4D}[\sigma]&=&-\frac{\epsilon^{AB}}{2\la \alpha_+\alpha_-\ra \la\alpha_+\beta\ra^2}\int_{\mathbb{E}^4}\on{tr}\(J_A^{\gamma_-}J_B^{\gamma_+}+iJ_A^{\gamma_-}R_\sigma J_B^{\gamma_-}\)\nn
&&+\frac{1}{6}\frac{1}{\la \alpha_+\alpha_-\ra \la\alpha_+\beta\ra^2} \int_{\mathbb{E}^4\times[0,1]} (\mu_{\alpha_+,\alpha_+}-\mu_{\alpha_-,\alpha_-})\wedge \on{tr} \(\tilde{\mathcal{J}}^3 \),
\eea
where $\mu_{\alpha_\pm,\alpha_\pm}\equiv\mathrm{d}^2x^{A'B'}{\alpha_\pm}_{A'}{\alpha_\pm}_{B'}$. This 4D theory is integrable in the sense that its equation of motion can be cast into the form of ASDYM. Without loss of generality we fix the normalization $\la \alpha_+\alpha_-\ra=1$ then perform the symmetry reduction according to \eqref{1.61} and \eqref{1.62}. The resulting  action of the 2D theory is
\bea \label{eta2d}
S_{2D}[\sigma]=\int_{\mathbb{E}^2} \on{tr} \( j_w(1+\eta R_\sigma)j_{\bar{w}}\)+\frac{k}{3}\int_{\mathbb{E}^2\times[0,1]} \on{tr}\({\tilde{j}^3}\),
\eea 
where we have dropped the overall factor $1/\la\alpha_+\beta\ra^2$ and defined 
\be
k=\la\alpha_+\mu \ra \la \alpha_+\hat{\mu}\ra- \la\alpha_-\mu \ra \la \alpha_-\hat{\mu}\ra, \hspace{2ex} \eta=i\la \gamma_-\mu\ra \la\gamma_-\hat{\mu}\ra.\ee
Unfortunately it is \textit{not} the action of the Yang-Baxter model, and moreover it has been shown in \cite{YBWZ} that the theory  \eqref{eta2d} is \textit{not} integrable. From the perspective of our construction, the non-integrability is due to the failure of the \textit{matching conditions} such that \eqref{eta2d}  can not be obtained from the 4DCS theory. It is a kind of puzzling that under symmetry reduction the 4D Lax connection can not be pushed forward to the Lax connection of a 2D integrable system. To understand the failure of the pushforward, let us contract the ASDYM with $\beta^{A'}\beta^{B'}$, $\beta^{A'}\hat{\beta}^{B'}$ and $\hat{\beta}^{A'}\hat{\beta}^{B'}$, and obtain
\bea 
&&\beta^{A'}\beta^{B'}\mathcal{E}_{A'B'}=0, \label{Con4}\\
&&\hat{\beta}^{A'}\hat{\beta}^{B'}\mathcal{E}_{A'B'}=\epsilon^{AB}\(\hat{\beta}^{A'}\p_{AA'}A_B-\hat{\beta}^{B'}\p_{BB'}A_A-[A_A,A_B]\)=0,\label{Con5}\\
&&\beta^{A'}\hat{\beta}^{B'}\mathcal{E}_{A'B'}=\epsilon^{AB}\beta^{A'}\p_{AA'}A_B=0.\label{Con6}
\eea 
As before, the symmetry reduction of \eqref{Con6} gives the equation of motion of the 2D theory \eqref{eta2d}
\be
\epsilon^{AB}\beta^{A'}\p_{AA'}A_B=-\la\beta\mu \ra \p_w[\hat{k}A]+\la\beta\hat{\mu}\ra\p_{\bar{w}}[kA]=\frac{1}{2\la \alpha_+ \beta\ra}\(\p_w \mathcal{K}_{\bar{w}}+\p_{\bar{w}}\mathcal{K}_w\)=0,\ee
with
\bea 
&&\mathcal{K}_{\bar{w}}=\(\la \gamma_- \mu\ra \la \gamma_+\hat{\mu}\ra+i\la\gamma_-\mu\ra \la \gamma_-\hat{\mu}\ra R_\sigma\)j_{\bar{w}}=(1+k+\eta R_{\sigma})j_{\bar{w}},\nn
&&\mathcal{K}_{\bar{w}}=\(-\la \gamma_- \hat{\mu}\ra \la \gamma_+{\mu}\ra-i\la\gamma_-\hat{\mu}\la \gamma_-{\mu}\ra R_\sigma\)j_{\bar{w}}=(1-k-\eta R_{\sigma})j_{{w}},\nonumber
\eea
where we have used the fact as $\la \gamma_-\beta\ra=0$, $\gamma_- \propto \beta$. Similarly the symmetry reduction of \eqref{Con5} leads to
\bea 
&& \p_w\mathcal{K}_{\bar{w}}- \p_{\bar{w}}\mathcal{K}_{{w}}+[\mathcal{K}_w,\mathcal{K}_{\bar{w}}]+x\(\p_w \mathcal{K}_{\bar{w}}+\p_{\bar{w}}\mathcal{K}_w\)=0,\eea
with
\bea
&&x=\la\alpha_+\beta\ra \(\la\hat{\beta}\mu\ra\la\gamma_-\hat{\mu}\ra+\la\hat{\beta}\hat{\mu}\ra\la\gamma_-{\mu}\ra\).
\eea 
However the flatness of $\mathcal{K}_{w(\bar{w})}$ is incompatible with the flatness of $j_{w(\bar{w})}$  when $\eta \neq 0$, as shown in \cite{YBWZ}.  Consequently, the symmetry reduction of $A_{AA'}'$ is not a 2D Lax connection anymore.

\subsection{Generalized $\lambda$--deformation}

The 2D generalized $\lambda$--deformed models have been successfully constructed from 4DCS in \cite{Bassi:2019aaf}. In this case, the 1-form $\omega$ has only pairs of simple poles. This suggests us to  consider the $(3,0)$--form $\Omega$ with four simple poles in 6DhCS
\bea 
  \label{Omega2}
    \Omega=\frac{\langle\mathrm{d}\pi\pi\rangle\wedge\mathrm{d}^2x^{A'B'}\pi_{A'}\pi_{B'}}{2\langle\pi\alpha_+\rangle\langle\pi\alpha_-\rangle\langle\pi\beta_-\rangle \la \pi\beta_+\ra},
\eea 
with the conditions
\bea \label{Con}
\la \alpha_+ \beta_-\ra \la\alpha_+\beta_+\ra=\la \alpha_- \beta_-\ra \la\alpha_-\beta_+\ra,\quad \la \alpha_+\beta_+\ra\la\alpha_-\beta_+\ra=\la \alpha_+\beta_-\ra\la\alpha_-\beta_-\ra.
\eea 
These conditions lead to
\be
\on{Res}_{\alpha_+}\Omega+\on{Res}_{\alpha_-}\Omega=\on{Res}_{\beta_+}\Omega+\on{Res}_{\beta_-}\Omega=0.\ee
Parameterizing the spinors as\footnote{One can choose a more general parameterization but it will not change the conclusion.}
\bea 
\alpha_+=(0,1),\quad \alpha_-=(1,0),\quad \beta_+=(b_1,b_2),\quad \beta_-=(b_3,b_4),
\eea 
we can solve the conditions \eqref{Con} by
\bea 
\beta_+=(a,b),\quad \beta_{-}=\pm (b,a).
\eea
Moreover we impose the requirement
\be
a\neq0,\quad b\neq0,\quad |a|\neq |b|
\ee
 such that the four poles do not collide. At the four simple poles we impose the boundary conditions
\bea\label{4poles} 
\hat{\mathcal{A}}\, |_{\alpha_+}=\hat{\mathcal{A}}\, |_{\alpha_-},\quad (R-i)\hat{\mathcal{A}}\, |_{\beta_+}=(R+i)\hat{\mathcal{A}}\, |_{\beta_-},
\eea 
and in order to remove gauge freedoms we fix 
\bea 
\hat{\sigma}\, |_{\alpha_+}=\hat{\sigma}\, |_{\beta_+}=\hat{\sigma}\, |_{\beta_-},\quad \hat{\sigma}\, |_{\alpha_-}=\sigma.
\eea 
With this choice of boundary condition \eqref{4poles}, one might expect to obtain a 4D analogue of generalized $\lambda$--deformed models \cite{Gen} which contain more than one deformation parameter. However we will show below that the resulting 4D theories actually  has only one deformation parameter.

Assuming $\hat{\mathcal{A}}=\pi^{A'}A_{AA'}$  and using \eqref{4dLax}, we arrive at the equations
\bea 
&&\D_\sigma A_A^{\alpha_+}=-J_A^{\alpha_-}+A_A^{\alpha_-},\quad (R-i)A_A^{\beta_+}=(R+i)A_A^{\beta_-},
\eea
from which one can solve
\bea 
&&A^{\alpha_+}_A=\frac{J_A^{\alpha_-}}{\delta N-\D_\sigma},\quad A^{\alpha_-}_A=\frac{J_A^{\alpha_-}}{1-\delta \D N^{-1}},\eea
with 
\be
\delta=\pm 1,\quad N=\frac{\eta R+1}{\eta R-1},\quad \eta=-i \frac{a-\delta b}{a+\delta b} .
\ee
Therefore we find
\bea 
&&\varepsilon^{AB}\on{tr}\( J^{\alpha_-}_A A_B^{\alpha_-} \) =\frac{\varepsilon^{AB}}{2}\on{tr}\( J_A^{\alpha_-}[(1-\delta \D_\sigma N^{-1})^{-1}-(1-\delta N^{-T}\D_{\sigma}^T)^{-1}] J_B^{\alpha_-}\)
\eea
The symmetry reduction along $H$ leads to 
\bea 
\lefteqn{\la \alpha_-\mu\ra \la \alpha_-\hat{\mu}\ra \on{tr} (J_{\bar{w}} [(1-\delta \D_\sigma N^{-1})^{-1}-(1-\delta N^{-T}\D_{\sigma}^T)^{-1}]J_w)} \nonumber \\
&&=\la \alpha_-\mu\ra \la \alpha_-\hat{\mu}\ra \on{tr} \( J_{\bar{w}} J_w+2 J_{\bar{w}} \frac{1}{\delta N \D_\sigma^T-1} J_w\),
\eea
which reproduces the action of the generalized  $\lambda$--deformed PCM \cite{Gen} with
\bea 
\lambda^{-1}=\delta N.
\eea 
However like the situation of $\lambda$--deformation, there is only \textit{one} free parameter.

\section{Coupled $\lambda$--deformation: an example with a fourth-order pole}
In the study of $\lambda$-deformation, we noticed that the failure in constructing 4D deformed model could due to the absence of zero in the $(3,0)$-form $\Omega$ and no nontrivial defect inserted. 
It would be interesting to consider the $(3,0)$-form $\Omega$ with zeros. 
The simplest $(3,0)$-form with zeros and even numbers of simple poles is 
\bea 
 \label{Omega4}
    \Omega=\frac{\langle\mathrm{d}\pi\pi\rangle\wedge\mathrm{d}^2x^{A'B'}\pi_{A'}\pi_{B'}}{2\langle\pi\alpha_+\rangle\langle\pi\alpha_-\rangle\langle\pi\beta\rangle^2}\frac{\la \pi \mu_+\ra \la \pi \mu_-\ra}{\la \pi\beta\ra^2}.
\eea 
In this case, there is a fourth-order pole. In the next section, we will consider another $(3,0)$-form $\Omega$ with zeros and two double poles. 
As before we require $\on{Res}_{\alpha_+}\Omega+\on{Res}_{\alpha_-}\Omega=0$, i.e.
\bea
\frac{\la \alpha_+\mu_+\ra \la \alpha_+\mu_-\ra}{\la \alpha_+\beta \ra^4}= \frac{\la \alpha_-\mu_+\ra \la \alpha_-\mu_-\ra}{\la \alpha_-\beta\ra^4}.
\eea 
To avoid the singularity at the position of the fourth-order pole in the action \eqref{equ:1.1}, the gauge connection there should be proportional to $\langle\pi\beta\rangle^2$. Therefore the proper boundary conditions are\footnote{Here we only focus on one of the possible Lagrangian subalgebras.}
\bea 
    \label{f.2}
    \hat{\mathcal{A}}|_{\pi=\beta}\propto\langle\pi\beta\rangle^2,\qquad \hat{\mathcal{A}}_{A}|_{\pi=\alpha_+}=\hat{\mathcal{A}}_{A}|_{\pi=\alpha_-}. 
\eea 
Because the gauge transformation \eqref{4dLax} should be compatible with this boundary condition we can not use the \textit{archipelago condition} to set $\hat{\sigma}_\beta=\on{id}$. Instead it was shown in \cite{Skiner} that the proper \textit{archipelago condition} at the fourth-order pole is $\hat{\sigma}\sim \operatorname{exp}\left(-\frac{\langle\pi\beta\rangle\langle\hat{\pi}\beta\rangle}{||\pi||^2}\phi\right) $, where $\phi$ is a regular function valued in $\mathfrak{g}$. At the two simple poles we can remove the residue gauge freedoms as before.  Therefore we have
\begin{align}
    \label{f.1}
    \hat{\sigma}|_{V_{\alpha_+}}=\sigma,\qquad \hat{\sigma}|_{V_{\alpha_-}}=\operatorname{id},\qquad \hat{\sigma}|_{V_\beta}=\operatorname{exp}\left(-\frac{\langle\pi\beta\rangle\langle\hat{\pi}\beta\rangle}{||\pi||^2}\phi\right)
\end{align}
where $V_{\alpha_+}, V_{\alpha_-}, V_\beta$ denote small neighborhoods of $\alpha_+, \alpha_-, \beta$, respectively. 

Combining \eqref{f.1} with \eqref{f.2} and considering analytic structure of the $(3,0)$-form $\Omega$, we expect that the 4D connection $\bar{\mathcal{A}}'$ should be proportional to $\langle\pi\beta\rangle$ and have two simple poles at $\mu_+$ and $\mu_-$. Thus we make the following ansatz 
\begin{align}
    \label{f.3}
    \hat{\mathcal{A}}'_A=\langle\pi\beta\rangle\left(\hat{\kappa}_A\frac{\langle\pi\mathcal{A}_{\kappa}\rangle}{\langle\pi\mu_+\rangle}-\kappa_{A}\frac{\langle\pi\mathcal{A}_{\hat{\kappa}}\rangle}{\langle\pi\mu_-\rangle}\right),
\end{align}
where $\kappa_A$ is a spinor of norm one, i.e., $||\kappa||^2=1$. Note that ${\mathcal{A}_{\kappa}}_{A'}$ and ${\mathcal{A}_{\hat{\kappa}}}_{A'}$ do not depend on $\pi^{A'}$ here. For later convenience, we expand ${\mathcal{A}_{\kappa}}_{A'}$ and ${\mathcal{A}_{\hat{\kappa}}}_{A'}$ in the following form
\begin{align}
    \label{f.4}
    {\mathcal{A}_{\kappa}}_{A'}\equiv\beta_{A'}A^{(\beta)}_{\kappa}+\hat{\beta}_{A'}A_{\kappa}^{(\hat{\beta})},\qquad {\mathcal{A}_{\hat{\kappa}}}_{A'}\equiv\beta_{A'}A^{(\beta)}_{\hat{\kappa}}+\hat{\beta}_{A'}A^{(\hat{\beta})}_{\hat{\kappa}}. 
\end{align}
The first  boundary condition in \eqref{f.2} is equivalent to 
\begin{align}
    \label{f.5}
    \hat{\kappa}_A\frac{\langle\beta\mathcal{A}_{\kappa}\rangle}{\langle\beta\mu_+\rangle}-\kappa_{A}\frac{\langle\beta\mathcal{A}_{\hat{\kappa}}\rangle}{\langle\beta\mu_-\rangle}+\beta^{A'}\partial_{AA'}\phi=0. 
\end{align}
Putting the decomposition \eqref{f.4} into these equations and solving the resulting equations, we get 
\begin{align}
    \label{f.6}
    A_{\kappa}^{(\hat{\beta})}=-\langle\beta\mu_+\rangle\kappa^{A}\beta^{A'}\partial_{AA'}\phi,\qquad A_{\hat{\kappa}}^{(\hat{\beta})}=-\langle\beta\mu_-\rangle\hat{\kappa}^A \beta^{A'}\partial_{AA'}\phi. 
\end{align}
Substituting the ansatz \eqref{f.3} in the second boundary equation in \eqref{f.2} gives
\begin{align}
    \label{f.7}
    \sigma^{-1}\alpha_+^{A'}\partial_{AA'}\sigma+\langle\alpha_+\beta\rangle\operatorname{Ad}_{\sigma^{-1}}\left(\hat{\kappa}_{A}\frac{\langle\alpha_+\mathcal{A}_{\kappa}\rangle}{\langle\alpha_+\mu_+\rangle}-\kappa_{A}\frac{\langle\alpha_+\mathcal{A}_{\hat{\kappa}}\rangle}{\langle\alpha_+\mu_-\rangle}\right)=\langle\alpha_-\beta\rangle\left(\hat{\kappa}_{A}\frac{\langle\alpha_-\mathcal{A}_{\kappa}\rangle}{\langle\alpha_-\mu_+\rangle}-\kappa_{A}\frac{\langle\alpha_-\mathcal{A}_{\hat{\kappa}}\rangle}{\langle\alpha_-\mu_-\rangle}\right). 
\end{align}
By contracting it with $\kappa^{A}$ and $\hat{\kappa}^{A}$, we find $A^{(\beta)}_{\kappa}$ and $A^{(\beta)}_{\hat{\kappa}}$ 
\begin{align}
    \label{f.8}
    A^{(\beta)}_{\kappa}&=\frac{\langle\alpha_+\mu_+\rangle}{\langle\alpha_+\beta\rangle^2}\frac{1}{1-\lambda\operatorname{Ad}_{\sigma}}\left(-\kappa^{A}\alpha_+^{A'}\partial_{AA'}\sigma\sigma^{-1}-\frac{\langle\alpha_+\beta\rangle\langle\alpha_+\hat{\beta}\rangle}{\langle\alpha_+\mu_+\rangle}A_{\kappa}^{(\hat{\beta})}+\frac{\langle\alpha_-\beta\rangle\langle\alpha_-\hat{\beta}\rangle}{\langle\alpha_-\mu_+\rangle}\operatorname{Ad}_{\sigma}A_{\kappa}^{(\hat{\beta})}\right),\\
    \label{f.9}
    A^{(\beta)}_{\hat{\kappa}}&=\frac{\langle\alpha_+\mu_-\rangle}{\langle\alpha_+\beta\rangle^2}\frac{1}{1-\lambda^{-1}\operatorname{Ad}_{\sigma}}\left(-\hat{\kappa}^{A}\alpha_+^{A'}\partial_{AA'}\sigma\sigma^{-1}-\frac{\langle\alpha_+\beta\rangle\langle\alpha_+\hat{\beta}\rangle}{\langle\alpha_+\mu_-\rangle}A^{(\hat{\beta})}_{\hat{\kappa}}+\frac{\langle\alpha_-\beta\rangle\langle\alpha_-\hat{\beta}\rangle}{\langle\alpha_-\mu_-\rangle}\operatorname{Ad}_{\sigma}A^{(\hat{\beta})}_{\hat{\kappa}}\right),
\end{align}
where \be\label{4polelambda}
\lambda\equiv\frac{\langle\alpha_-\beta\rangle^2\langle\alpha_+\mu_+\rangle}{\langle\alpha_+\beta\rangle^2\langle\alpha_-\mu_+\rangle}=\frac{\langle\alpha_+\beta\rangle^2\langle\alpha_-\mu_-\rangle}{\langle\alpha_-\beta\rangle^2\langle\alpha_+\mu_-\rangle}.\ee
 The action $S_{\Omega}[\bar{\mathcal{A}}]$ in this case get contributions from the poles of $\Omega$ both at $\alpha_+$ and $\beta$, i.e. 
\begin{align}
    \label{f.10}
    S_{\Omega}[\bar{\mathcal{A}}]=&-\frac{1}{2\pi i} \lim _{\epsilon \rightarrow 0} \oint_{S_{\alpha_+, \epsilon}^{1}}\left[\frac{\langle\mathrm{d} \pi \pi\rangle\langle\pi\mu_+\rangle\langle\pi\mu_-\rangle}{\langle\pi\alpha_+\rangle\langle\pi\alpha_-\rangle\langle\pi\beta\rangle^4} \int_{\mathbb{E}^{4}} \operatorname{vol}_4\varepsilon^{AB} \operatorname{tr}(\hat{\mathcal{J}}_A  \hat{\mathcal{A}}'_B)\right]\notag\\
    &-\frac{1}{2\pi i} \lim _{\epsilon \rightarrow 0} \oint_{S_{\beta, \epsilon}^{1}}\left[\frac{\langle\mathrm{d} \pi \pi\rangle\langle\pi\mu_+\rangle\langle\pi\mu_-\rangle}{\langle\pi\alpha_+\rangle\langle\pi\alpha_-\rangle\langle\pi\beta\rangle^4} \int_{\mathbb{E}^{4}} \operatorname{vol}_4\varepsilon^{AB} \operatorname{tr}(\hat{\mathcal{J}}_A  \hat{\mathcal{A}}'_B)\right]\notag\\
    &+\frac{1}{12\pi i}\int_{\mathbb{PT}}\frac{\langle\mathrm{d}\pi \pi\rangle\wedge\mathrm{d}^{2}x^{A'B'}\pi_{A'}\pi_{B'}\langle\pi\mu_+\rangle\langle\pi\mu_-\rangle}{\langle\pi\alpha_+\rangle\langle\pi\alpha_-\rangle\langle\pi\beta\rangle^4}\operatorname{tr}\left(\bar{\mathcal{J}^3}\right). 
\end{align}
The contribution from the pole at $\alpha_+$ is simply
\iffalse
\begin{align}
    \label{f.11}
    -\frac{\langle\alpha_+\mu_+\rangle\langle\alpha_+\mu_-\rangle}{\langle\alpha_+\alpha_-\rangle\langle\alpha_+\beta\rangle^4}\operatorname{tr}&\left(-\kappa^A\alpha_+^{A'}\partial_{AA'}\sigma\sigma^{-1}\frac{\langle\alpha_+\beta\rangle}{\langle\alpha_+\mu_-\rangle}\left(\langle\alpha_+\beta\rangle A^{(\beta)}_{\hat{\kappa}}+\langle\alpha_+\hat{\beta}\rangle A^{(\hat{\beta})}_{\hat{\kappa}}\right)\right.\notag\\
    &\left.+\hat{\kappa}^A \alpha_+^{A'}\partial_{AA'}\sigma\sigma^{-1}\frac{\langle\alpha_+\beta\rangle}{\langle\alpha_+\mu_+\rangle}\left(\langle\alpha_+\beta\rangle A^{(\beta)}_{\kappa}+\langle\alpha_+\hat{\beta}\rangle A^{(\hat{\beta})}_{\kappa}\right)\right)
\end{align}
By putting in solutions \eqref{f.8} and \eqref{f.9} into \eqref{f.11} we get
\fi
\begin{align}
    \label{f.12}
    -\frac{\langle\alpha_+\mu_+\rangle\langle\alpha_+\mu_-\rangle}{\langle\alpha_+\alpha_-\rangle\langle\alpha_+\beta\rangle^4}&\left[\operatorname{tr}\left(\hat{\kappa}^{A}\alpha_+^{A'}\partial_{AA'}\sigma\sigma^{-1}\left(1-\frac{2}{1-\lambda\operatorname{Ad}_{\sigma}}\right)\kappa^B \alpha_+^{B'}\partial_{BB'}\sigma\sigma^{-1}\right)\right.\notag\\
    &-\frac{\langle\alpha_-\beta\rangle\langle\alpha_+\alpha_-\rangle}{\langle\alpha_+\beta\rangle\langle\alpha_-\mu_-\rangle}\operatorname{tr}\left(\kappa^A \alpha_+^{A'}\partial_{AA'}\sigma\sigma^{-1}\frac{\operatorname{Ad}_{\sigma}}{1-\lambda^{-1}\operatorname{Ad}_{\sigma}}A^{(\hat{\beta})}_{\hat{\kappa}}\right)\notag\\
    &\left.+\frac{\langle\alpha_-\beta\rangle\langle\alpha_+\alpha_-\rangle}{\langle\alpha_+\beta\rangle\langle\alpha_-\mu_+\rangle}\operatorname{tr}\left(\hat{\kappa}^A \alpha_+^{A'}\partial_{AA'}\sigma\sigma^{-1}\frac{\operatorname{Ad}_{\sigma}}{1-\lambda\operatorname{Ad}_{\sigma}}A^{(\hat{\beta})}_{\kappa}\right)\right]
\end{align}
with $A_{\kappa}^{(\hat{\beta})}$ and $A_{\hat{\kappa}}^{(\hat{\beta})}$ being given in \eqref{f.6}. On the other hand the contribution from the pole at $\beta$ is
\begin{align}
    \label{f.13}
    \frac{\langle\beta\mu_+\rangle\langle\beta\mu_-\rangle}{\langle\beta\alpha_+\rangle\langle\beta\alpha_-\rangle}&\operatorname{tr}\left[-\frac{1}{2}\varepsilon^{AB}\varepsilon^{A'B'}\partial_{AA'}\phi\partial_{BB'}\phi+\left(\frac{\langle\hat{\beta}\mu_+\rangle}{\langle\beta\mu_+\rangle}\hat{\kappa}^A\kappa^B-\frac{\langle\hat{\beta}\mu_-\rangle}{\langle\beta\mu_-\rangle}\kappa^A \hat{\kappa}^{B}\right)\beta^{A'}\beta^{B'}\partial_{AA'}\phi\partial_{BB'}\phi\right.\notag\\
    &\left.+\kappa^A\beta^{A'}\partial_{AA'}\phi\frac{A^{(\beta)}_{\hat{\kappa}}}{\langle\beta\mu_-\rangle}-\hat{\kappa}^A\beta^{A'}\partial_{AA'}\phi\frac{A^{(\beta)}_{\kappa}}{\langle\beta\mu_+\rangle}\right].
\end{align}
The topological term can be obtained as before so we do not repeat the analysis here. By adding \eqref{f.12} and \eqref{f.13} up, we get the action of the 4D theory. Note that the first line of \eqref{f.12} is very similar to $\lambda$-deformed principal chiral model in 2D (one can further check it by doing symmetry reduction). However, we also get some extra terms which couple the $\sigma$ field to a new scalar field $\phi$. Thus this model should not be viewed as 4D analogue of $\lambda$-deformed principal chiral model. It has not been fully studied in the literature. 

One can also do the symmetry reduction directly on the twistor space which yields a 4DCS theory on $\mathbb{E}^2\times\mathbb{CP}^1$ with 1-form
\bea\label{4poles1form}
\omega=\frac{\la \pi d\pi\ra \la  \pi \mu_+\ra\la \pi\mu_-\ra\la\pi \mu\ra\la \pi\hat{\mu}\ra}{\la\pi \alpha_+\ra\la\pi \alpha_-\ra \la\pi\beta\ra^4 }.
\eea 
The second boundary condition in \eqref{f.2} will not be compatible with the \textit{matching condition}. In order to proceed we impose the new condition and boundary condition
\bea 
&&\frac{\la\alpha_+\mu_+\ra\la\alpha_+\mu_-\ra\la\alpha_+\mu\ra\la\alpha_+\hat{\mu}\ra}{\la\alpha_+\beta\ra^4}=\frac{\la\alpha_-\mu_+\ra\la\alpha_-\mu_-\ra\la\alpha_-\mu\ra\la\alpha_-\hat{\mu}\ra}{\la\alpha_-\beta\ra^4}, \label{4polecon1}\\
&&\bar{A}_{w,\bar{w}}|_{\pi=\alpha_+}=\bar{A}_{w,\bar{w}}|_{\pi=\alpha_-}. \label{4polebdy}
\eea 
Then we can again fix the gauge as \eqref{f.1}. Considering that there are four zeros in the 1-form \eqref{4poles1form}, the 2D Lax connection should be in the form
\bea\label{4polesLax}
&&\mathcal{L}_w=\la\pi\beta\ra \(  \frac{\la\beta\mu_+\ra}{\la \pi\mu_+\ra}U_w+\frac{\la \beta\mu\ra}{\la \pi \mu\ra}V_w \),\\
&&\mathcal{L}_{\bar{w}}=\la\pi\beta\ra \(  \frac{\la\beta\mu_-\ra}{\la \pi\mu_-\ra}U_{\bar{w}}+\frac{\la \beta\hat{\mu}\ra}{\la \pi \hat{\mu}\ra}V_{\bar{w}} \),
\eea 
where $U$ and $V$ are regular function on $\mathbb{E}^2$. Thanks to the relation \eqref{equ:1.27} the boundary conditions of the gauge fields at the fourth-order pole and at the two simple poles are equivalent to 
\bea 
&&U_w+V_w=-\p_w \phi,\quad U_{\bar{w}}+V_{\bar{w}}=-\p_{\bar{w}} \phi,\\
&&\p_w \sigma\sigma^{-1}+\la \alpha_+\beta\ra \(\frac{\la\beta\mu_+\ra}{\la\alpha_+\mu_+\ra}U_w+\frac{\la\beta\mu\ra}{\la\alpha_+\mu\ra}V_w\)=\on{Ad}_\sigma \la\alpha_-\beta\ra \(\frac{\la\beta\mu_+\ra}{\la\alpha_-\mu_+\ra}U_w+\frac{\la\beta\mu\ra}{\la\alpha_-\mu\ra}V_w\),\nn
&&\p_{\bar{w}} \sigma\sigma^{-1}+\la \alpha_+\beta\ra \(\frac{\la\beta\mu_-\ra}{\la\alpha_+\mu_-\ra}U_{\bar{w}}+\frac{\la\beta\hat{\mu}\ra}{\la\alpha_+\hat{\mu}\ra}V_{\bar{w}}\)=\on{Ad}_\sigma \la\alpha_-\beta\ra \(\frac{\la\beta\mu_-\ra}{\la\alpha_-\mu_-\ra}U_{\bar{w}}+\frac{\la\beta\hat{\mu}\ra}{\la\alpha_-\hat{\mu}\ra}V_{\bar{w}}\),\nn
\eea 
from which one can solve 
\bea 
&&U_w=\frac{j_w+D_w \p_w\phi}{N_w (1-\lambda \on{Ad}_\sigma)},\quad U_{\bar{w}}=\frac{j_{\bar{w}}+D_{\bar{w}} \p_{\bar{w}}\phi}{N_{\bar{w}} (1-\lambda^{-1} \on{Ad}_\sigma)}
\eea 
with
\bea 
&&\lambda\equiv\frac{\langle\alpha_-\beta\rangle^2\langle\alpha_+\mu_+\rangle \la\alpha_+\mu\ra}{\langle\alpha_+\beta\rangle^2\langle\alpha_-\mu_+\rangle\la\alpha_-\mu\ra}=\frac{\langle\alpha_+\beta\rangle^2\langle\alpha_-\mu_-\rangle\la\alpha_-\hat{\mu}\ra}{\langle\alpha_-\beta\rangle^2\langle\alpha_+\mu_-\rangle\la\alpha_+\hat{\mu}\ra},\label{4poleslambda}\\
&&N_w=\la \alpha_+\beta\ra \(  \frac{\la\beta\mu_+\ra}{\la\alpha_+\mu_+\ra} - \frac{\la\beta\mu\ra}{\la\alpha_+\mu\ra} \),\quad N_{\bar{w}}=\la \alpha_+\beta\ra \(  \frac{\la\beta\mu_-\ra}{\la\alpha_+\mu_-\ra} - \frac{\la\beta\hat{\mu}\ra}{\la\alpha_+\hat{\mu}\ra} \),\\
&&D_{w}=\frac{\la\alpha_+\beta\ra\la\beta\mu\ra}{\la\alpha_+\mu\ra}-\frac{\la \alpha_-\beta\ra \la \beta\mu\ra}{\la\alpha_-\mu\ra}\on{Ad}_\sigma,\quad D_{\bar{w}}=\frac{\la\alpha_+\beta\ra\la\beta\hat{\mu}\ra}{\la\alpha_+\hat{\mu}\ra}-\frac{\la \alpha_-\beta\ra \la \beta\hat{\mu}\ra}{\la\alpha_-\hat{\mu}\ra}\on{Ad}_\sigma.
\eea 
Note that the definition of $\lambda$ here is different from \eqref{4polelambda}.
To derive the 2D action one only needs to substitute the 2D Lax connection to the action of 4DCS and integrate over $\mathbb{CP}^1$ with residue theorem\footnote{It is also straightforward but tedious to obtain the topological term as before so we will not repeat the analysis here}. The contribution to the Lagrangian density from the pole at $\alpha_+$ is
\begin{align}
\frac{\la \alpha_+\mu_+\ra\la\alpha_+ \mu_-\ra\la \alpha_+\mu\ra\la \alpha_+\mu\ra}{\la \alpha_+\alpha_-\ra \la\alpha_+\beta\ra^4} &\{ \on{tr}\( j_w(1-\frac{2}{1-\lambda \on{Ad}_{\sigma^{-1}} } )j_{\bar{w}}\)\notag\\
-&\frac{\la\beta\mu\ra}{\la\alpha_-\mu\ra \la\beta\mu_+\ra }\frac{\la\alpha_-\beta\ra\la\alpha_+\alpha_-\ra}{\la\alpha_+\beta \ra\la\alpha_-\mu_+\ra} \on{tr}\( j_{\bar{w}}\frac{\on{Ad}_\sigma}{1-\lambda \on{Ad}_\sigma}\p_w \phi\)\nn
+&\frac{\la\beta\hat{\mu}\ra}{\la \alpha_-\hat{\mu}\ra\la \beta \mu_-\ra}\frac{\la \alpha_-\beta\ra \la\alpha_+\alpha_-\ra}{\la\alpha_+\beta\ra\la\alpha_-\mu_-\ra}\on{tr}\(j_w \frac{\on{Ad}_\sigma}{1-\lambda^{-1}\on{Ad}_\sigma}\p_{\bar{w}}\phi\)
\}.
\end{align}
Similarly the contribution form the pole at $\beta$ is
\bea 
\frac{\la\beta \mu_+\ra\la \beta\mu_-\ra\la\beta\mu\ra \la \beta\hat{\mu}\ra}{\la \beta\alpha_+\ra \la \beta\alpha_-\ra}\on{tr}\( \p_w\phi[\frac{\la\hat{\beta}\mu_+\ra}{\la\beta\mu_+\ra^2}U_w+\frac{\la\hat{\beta}\mu\ra}{\la\beta{\mu}\ra^2}V_w]-\p_{\bar{w}}\phi[\frac{\la\hat{\beta}\mu_-\ra}{\la\beta\mu_-\ra^2}U_{\bar{w}}+\frac{\la\hat{\beta}\hat{\mu}\ra}{\la\beta\hat{\mu}\ra^2}V_{\hat{w}}]\right).\nn
\eea 
As expected the 2D model describes a $\lambda$--deformed model coupled with an additional field $\phi$. It would be very interesting to understand how to construct this coupled integrable model from only 2D point of view.

\section{$\eta$--deformation from the trigonometric description}
In  section \ref{4.1} we have shown that we can not construct the $\eta$--deformation by solving along fibre first from 6DhCS due to the fact that there is no zero in the $(3,0)$-form $\Omega$ and no non-trivial defect is inserted. In this section we will show that the $\eta$--deformation can be constructed if one considers its trigonometric description. The $\eta$--deformation in the trigonometric description has been studied from the point of view of 4DCS in \cite{Fukushima:2020kta}. An important observation made in \cite{Fukushima:2020kta} is that  in the trigonometric description the numbers of zeros and double poles of the corresponding 1-form $\omega$ are doubled comparing with the ones in the rational description such that the resulting 2D field theory has an additional $\mathbb{Z}_2$ symmetry,  and different types of Yang-Baxter models can be obtained by gauging the $\mathbb{Z}_2$ symmetry differently. The constructions in this section can be summarized in Fig. 4.
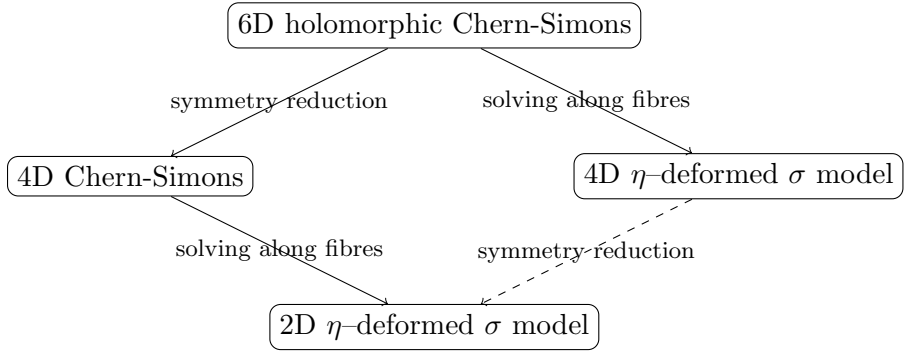
\begin{figure}[h]
\label{Fig4}
\begin{center}
\begin{tikzpicture}[scale=1]
\path (0,0) node[rectangle,
           rounded corners,draw](2D)  {2D $\eta$--deformed $\sigma$ model}
(0,4) node[rectangle,
           rounded corners,draw] (6D) {6D holomorphic Chern-Simons}
(-4,2) node[rectangle,
           rounded corners,draw] (4DCS) {4D Chern-Simons}
(4,2) node[rectangle,
           rounded corners,draw] (4DI) {4D $\eta$--deformed $\sigma$ model};
\draw[->] (6D) -- (4DCS)
 node[pos=0.5]{{\footnotesize symmetry reduction}};
\draw[->] (6D) -- (4DI)
node[pos=0.5]{{\footnotesize solving along fibres}};
\draw[->] (4DCS) -- (2D)
node[pos=0.5]{{\footnotesize solving along fibres}};
\draw[dashed,->] (4DI) -- (2D)
node[pos=0.5]{{\footnotesize symmetry reduction}};
\end{tikzpicture}
\caption{A summary of the construction of $\eta$--deformation in the trigonometric description.}
\end{center}
\end{figure}

From 6DhCS point of view, if we want to double the numbers of the double poles  we also need to introduce two zeros so the $(3,0)$-form $\Omega$ becomes
\begin{align}
    \label{1.4}
    \Omega=\frac{\langle\mathrm{d}\pi\pi\rangle\wedge\mathrm{d}^2x^{A'B'}\pi_{A'}\pi_{B'}\langle\pi\mu_{+}\rangle\langle\pi\mu_{-}\rangle}{2\langle\pi\alpha_+\rangle\langle\pi\alpha_-\rangle\langle\pi\beta\rangle^2\langle\pi\hat{\beta}\rangle^2}. 
\end{align}
At the two simple poles we still require that the residues are opposite which is equivalent to set
\begin{align}
    \label{c.1}
    \frac{\langle\alpha_{+}\mu_{+}\rangle\langle\alpha_{+}\mu_{-}\rangle}{\langle\alpha_{+}\beta\rangle^2\langle\alpha_{+}\hat{\beta}\rangle^2}=\frac{\langle\alpha_{-}\mu_{+}\rangle\langle\alpha_{-}\mu_{-}\rangle}{\langle\alpha_{-}\beta\rangle^2\langle\alpha_{-}\hat{\beta}\rangle^2}.
\end{align}

\subsection{Solving along fibres for holomorphic Chern-Simons theory}
Let us first  follow the right route  in Fig. 4. Similar to \eqref{Bdy2}, the proper boundary conditions should be
\begin{align}
    \label{c.6}
    \bar{\mathcal{A}}|_{\pi=\beta}=&0=\bar{\mathcal{A}}|_{\pi=\hat{\beta}}\\
    \label{c.7}
    (R-i)\hat{\mathcal{A}}_{A}|_{\pi=\alpha_{+}}=&(R+i)\hat{\mathcal{A}}_{A}|_{\pi=\alpha_{-}},\quad A=1,2.
\end{align}
Recall that the 4D Lax connection $\hat{A}'$ is related to the 6D gauge connection through the formal gauge transformation
\begin{align}
    \label{1.50}
    \bar{\mathcal{A}}=\hat{\sigma}^{-1}\bar{\partial}\hat{\sigma}+\hat{\sigma}^{-1}\bar{\mathcal{A}}'\hat{\sigma}, 
\end{align}
where $\bar{\mathcal{A}}'=\hat{e}^{A}\hat{\mathcal{A}}'_A$ does not have components in $\mathbb{CP}^1$, and  $\tilde{\sigma}$ satisfies the \textit{archipelago conditions} so that
\bea \label{Fix}
\hat{\sigma}|_{\alpha_+}=\sigma_1,\quad \hat{\sigma}|_{\alpha_-}=\sigma_2,\quad \hat{\sigma}|_{\beta}=\sigma,\quad \hat{\sigma}|_{\hat{\beta}}=\hat{\sigma}.
\eea 
We may use the gauge freedom $\hat{\sigma}\mapsto \hat{\sigma}g^{-1}$, where $g=\exp(i x),\,x\in \mathfrak{g}_R$ to fix $\sigma_1=\sigma_2=\sigma'$, then we use the gauge freedom $\hat{\sigma}\mapsto h \hat{\sigma} $, where $h:\mathbb{E}^4\rightarrow G$  to set $\sigma'=\on{id}$. The gauge connections have to have poles at the simple zeros of $\Omega$, so we make the following ansatz
\begin{align}
    \label{6}
    \bar{\mathcal{A}}'_A&=-\kappa_A[\hat{\kappa}\bar{\mathcal{A}}']+\hat{\kappa}_A[\kappa\bar{\mathcal{A}}'],\\
    \label{7}
    [\kappa\bar{\mathcal{A}}']&=\frac{\langle\pi\beta\rangle\langle\pi\hat{\beta}\rangle}{\langle\pi\mu_-\rangle}U^{(1)}_{\kappa}+\langle\pi\beta\rangle U_{\kappa}+\langle\pi\hat{\beta}\rangle \hat{U}_{\kappa},\\
    \label{c.8}
    [\hat{\kappa}\bar{\mathcal{A}}']&=\frac{\langle\pi\beta\rangle\langle\pi\hat{\beta}\rangle}{\langle\pi\mu_+\rangle}U^{(1)}_{\hat{\kappa}}+\langle\pi\beta\rangle U_{\hat{\kappa}}+\langle\pi\hat{\beta}\rangle \hat{U}_{\hat{\kappa}}
\end{align}
where $U^{(1)}_{\kappa,\hat{\kappa}}, U_{\kappa,\hat{\kappa}}, \hat{U}_{\kappa,\hat{\kappa}}$ do not depend on $\mathbb{CP}^1$. It is crucial that the gauge connections have weight $(1,0)$ even though it is not linear in $\pi$ anymore. Substituting the ansatz into the boundary conditions \eqref{c.6} gives
\begin{align}
    \label{c.9}
    \hat{U}_{\kappa}&=-\kappa^A \beta^{A'}\partial_{AA'}\sigma \sigma^{-1}\equiv J_{\kappa}^{\beta}, \quad \hat{U}_{\hat{\kappa}}=-\hat{\kappa}\beta^{A'}\partial_{AA'}\sigma \sigma^{-1}\equiv J_{\hat{\kappa}}^{\beta},\\
    \label{c.10}
    U_{\kappa}&=\kappa^A \hat{\beta}^{A'}\partial_{AA'}\hat{\sigma} \hat{\sigma}^{-1}\equiv -\hat{J}_{\kappa}^{\hat{\beta}}, \quad U_{\hat{\kappa}}=\hat{\kappa}^A\hat{\beta}^{A'}\partial_{AA'}\hat{\sigma} \hat{\sigma}^{-1}\equiv -\hat{J}_{\hat{\kappa}}^{\hat{\beta}}, \\
    \label{c.11}
    &(R-i)\left(\frac{\langle\alpha_+\beta\rangle\langle\alpha_+\hat{\beta}\rangle}{\langle\alpha_+\mu_-\rangle}U^{(1)}_{\kappa}+\langle\alpha_+\beta\rangle U_{\kappa}+\langle\alpha_+\hat{\beta}\rangle \hat{U}_{\kappa}\right) \notag\\
    &\hspace{5ex}=(R+i)\left(\frac{\langle\alpha_-\beta\rangle\langle\alpha_-\hat{\beta}\rangle}{\langle\alpha_-\mu_-\rangle}U^{(1)}_{\kappa}+\langle\alpha_-\beta\rangle U_{\kappa}+\langle\alpha_-\hat{\beta}\rangle \hat{U}_{\kappa}\right), \\
    \label{c.12}
    &(R-i)\left(\frac{\langle\alpha_+\beta\rangle\langle\alpha_+\hat{\beta}\rangle}{\langle\alpha_+\mu_+\rangle}U^{(1)}_{\hat{\kappa}}+\langle\alpha_+\beta\rangle U_{\hat{\kappa}}+\langle\alpha_+\hat{\beta}\rangle\hat{U}_{\hat{\kappa}}\right) \notag\\
   &\hspace{5ex}=(R+i)\left(\frac{\langle\alpha_-\beta\rangle\langle\alpha_-\hat{\beta}\rangle}{\langle\alpha_-\mu_+\rangle}U^{(1)}_{\hat{\kappa}}+\langle\alpha_-\beta\rangle U_{\hat{\kappa}}+\langle\alpha_-\hat{\beta}\rangle \hat{U}_{\hat{\kappa}}\right),
\end{align}
which are solved by
\begin{align}
    \label{c.13}
    U_{\kappa}^{(1)}&=-(\eta-i) \frac{\langle\alpha_{-} \mu_{-}\rangle}{2\langle\alpha_{-} \hat{\beta}\rangle\langle\alpha_{-} \beta\rangle} \frac{\left(\gamma_{-} R+i \gamma_{+}\right) U_{\kappa}+\left(\hat{\gamma}_{-} R+i \hat{\gamma}_{+}\right) \hat{U}_{\kappa}}{\eta R+1},\\
    \label{c.14}
    U_{\hat{\kappa}}^{(1)}&=(\eta-i)\frac{\langle\alpha_{+} \mu_{+}\rangle}{2\langle\alpha_{+} \hat{\beta}\rangle\langle\alpha_{+} \beta\rangle}\frac{\left(\gamma_{-} R+i \gamma_{+}\right) U_{\hat{\kappa}}+\left(\hat{\gamma}_{-} R+i \hat{\gamma}_{+}\right) \hat{U}_{\hat{\kappa}}}{\eta R-1},
\end{align} 
where we have defined the following quantities
\begin{align}
    \label{c.15}
    \gamma_{\pm}&=\left\langle\alpha_{-} \beta\right\rangle \pm\left\langle\alpha_{+} \beta\right\rangle, \quad \hat{\gamma}_{\pm}=\langle\alpha_{-} \hat{\beta}\rangle \pm\langle\alpha_{+} \hat{\beta}\rangle, \\
    \label{c.16}
    \eta&=-i \frac{1-\frac{\left\langle\alpha_{-} \beta\right\rangle\left\langle\alpha_{-} \hat{\beta}\right\rangle\left\langle\alpha_{+} \mu_{+}\right\rangle}{\left\langle\alpha_{+} \beta\right\rangle\left\langle\alpha_{+} \hat{\beta}\right\rangle\left\langle\alpha_{-} \mu_{+}\right\rangle}}{1+\frac{\left\langle\alpha_{-} \beta\right\rangle\left\langle\alpha_{-} \hat{\beta}\right\rangle\left\langle\alpha_{+} \mu_{+}\right\rangle}{\left\langle\alpha_{+} \beta\right\rangle\left\langle\alpha_{+} \hat{\beta}\right\rangle\left\langle\alpha_{-} \mu_{+}\right\rangle}}=-i \frac{1-\frac{\left\langle\alpha_{+} \beta\right\rangle\left\langle\alpha_{+} \hat{\beta}\right\rangle\left\langle\alpha_{-} \mu_{-}\right\rangle}{\left\langle\alpha_{-} \beta\right\rangle\left\langle\alpha_{-} \hat{\beta}\right\rangle\left\langle\alpha_{+} \mu_{-}\right\rangle}}{1+\frac{\left\langle\alpha_{+} \beta\right\rangle\left\langle\alpha_{+} \hat{\beta}\right\rangle\left\langle\alpha_{-} \mu_{-}\right\rangle}{\left\langle\alpha_{-} \beta\right\rangle\left\langle\alpha_{-} \hat{\beta}\right\rangle\left\langle\alpha_{+} \mu_{-}\right\rangle}}.
\end{align}
To derive the 4D action we need to substitute the 4D Lax connection we just obtained into the  action of 6DhCS
\begin{align}
    \label{1.56}
    S_{\Omega}[\bar{\mathcal{A}}]=&\frac{1}{2 \pi i} \int_{\mathbb{PT}} \Omega \wedge \mathrm{hCS}(\bar{\mathcal{A}})\notag\\
    =&\frac{1}{4\pi i}\sum_{z\in\mathfrak{p}} \lim _{\epsilon \rightarrow 0} \oint_{S_{z, \epsilon}^{1}}\left[\frac{\langle\mathrm{d} \pi \pi\rangle\langle\pi\mu_{+}\rangle\langle\pi\mu_{-}\rangle}{\langle\pi\alpha_+\rangle\langle\pi\alpha_-\rangle\langle\pi\beta\rangle^2\langle\pi\hat{\beta}\rangle^2} \int_{\mathbb{E}^{4}} \mathrm{d}^2 x^{A'B'}\pi_{A'}\pi_{B'}\wedge \operatorname{tr}(\bar{\mathcal{J}} \wedge \bar{\mathcal{A}}')\right]\notag\\
    &+\frac{1}{12\pi i}\int_{\mathbb{PT}}\frac{\langle\mathrm{d}\pi \pi\rangle\wedge\mathrm{d}^{2}x^{A'B'}\pi_{A'}\pi_{B'}\langle\pi\mu_{+}\rangle\langle\pi\mu_{-}\rangle}{\langle\pi\alpha_+\rangle\langle\pi\alpha_-\rangle\langle\pi\beta\rangle^2\langle\pi\hat{\beta}\rangle^2}\operatorname{tr}\left(\bar{\mathcal{J}^3}\right). 
\end{align}
Because $\tilde{\sigma}$ is not trivial at both $\beta$ and $\hat{\beta}$, \eqref{1.56} is given by
\begin{align}
    \label{1.57}
    &\frac{1}{4\pi i}\sum_{z\in\mathfrak{p}} \lim _{\epsilon \rightarrow 0} \oint_{S_{z, \epsilon}^{1}}\left[\frac{\langle\mathrm{d} \pi \pi\rangle\langle\pi\mu_{+}\rangle\langle\pi\mu_{-}\rangle}{\langle\pi\alpha_+\rangle\langle\pi\alpha_-\rangle\langle\pi\beta\rangle^2\langle\pi\hat{\beta}\rangle^2} \int_{\mathbb{E}^{4}} \mathrm{d}^2 x^{A'B'}\pi_{A'}\pi_{B'}\wedge \operatorname{tr}(\bar{\mathcal{J}} \wedge \bar{\mathcal{A}}')\right]\notag\\
    &\hspace{5ex}=-\frac{1}{2\pi i}\lim_{\epsilon\rightarrow 0} \oint_{S_{\beta, \epsilon}^{1}}\left[\frac{\langle\mathrm{d} \pi \pi\rangle\langle\pi\mu_{+}\rangle\langle\pi\mu_{-}\rangle}{\langle\pi\alpha_+\rangle\langle\pi\alpha_-\rangle\langle\pi\beta\rangle^2\langle\pi\hat{\beta}\rangle^2}\int_{\mathbb{E}^{4}}\operatorname{vol}_4  \varepsilon^{AB}\operatorname{tr}(\hat{\mathcal{J}}_A  \hat{\mathcal{A}}'_B)\right]\notag\\
    &\hspace{8ex}-\frac{1}{2\pi i}\lim_{\epsilon\rightarrow 0} \oint_{S_{\hat{\beta}, \epsilon}^{1}}\left[\frac{\langle\mathrm{d} \pi \pi\rangle\langle\pi\mu_{+}\rangle\langle\pi\mu_{-}\rangle}{\langle\pi\alpha_+\rangle\langle\pi\alpha_-\rangle\langle\pi\beta\rangle^2\langle\pi\hat{\beta}\rangle^2}\int_{\mathbb{E}^{4}}\operatorname{vol}_4  \varepsilon^{AB}\operatorname{tr}(\hat{\mathcal{J}}_A  \hat{\mathcal{A}}'_B)\right]. 
\end{align}
To evaluate this contour integral, we need to expand $\Omega \hat{\mathcal{J}}_A\hat{\mathcal{A}}_B'$ in powers of $\langle\pi\beta\rangle$ and $\langle\pi\hat{\beta}\rangle$. The expansion of  $\hat{\mathcal{J}}_A$ is
\begin{align}
    \label{c.17}
    \hat{\mathcal{J}}_A&\equiv-\pi^{A'}\partial_{AA'}\hat{\sigma}\hat{\sigma}^{-1}=-\langle\pi\hat{\beta}\rangle\beta^{A'}\partial_{AA'}\hat{\sigma}\hat{\sigma}^{-1}+\langle\pi\beta\rangle\hat{\beta}^{A'}\partial_{AA'}\hat{\sigma}\hat{\sigma}^{-1}.
\end{align}
Then we can easily find that the  the contribution from the terms proportional to $\langle\pi\beta\rangle$ is
\begin{align}
    \label{c.18}
    -\frac{1}{2\pi i}\varepsilon^{AB}\operatorname{tr}\left(\omega_1 J^{\beta}_{A} J^{\beta}_{B}-\omega_0 J^{\hat{\beta}}_{A} J^{\beta}_{B}-\omega_0 J^{\beta}_{A} \hat{J}^{\hat{\beta}}_{B}+\omega_0 J^{\beta}_{A}K_{B}\right),
\end{align}
where
\begin{align}
    &\frac{\left\langle\pi \mu_{-}\right\rangle\left\langle\pi \mu_{+}\right\rangle}{\left\langle\pi \alpha_{+}\right\rangle\left\langle\pi \alpha_{-}\right\rangle}=\omega_{0}+\omega_{1}\langle\pi \beta\rangle+\mathcal{O}\left(\langle\pi \beta\rangle^{2}\right),\notag\\
    &K_{A}=-\frac{\kappa_{A}}{\langle\beta\mu_{+}\rangle}U^{(1)}_{\hat{\kappa}}+\frac{\hat{\kappa}_{A}}{\langle\beta\mu_-\rangle}U^{(1)}_{\kappa}. \notag
\end{align}
And the contribution from the terms proportional to $\langle\pi\hat{\beta}\rangle$ is
\begin{align}
    \label{c.21}
    -\frac{1}{2\pi i}\varepsilon^{AB}\operatorname{tr}\left(-\hat{\omega}_1\hat{J}^{\hat{\beta}}_{A} \hat{J}^{\hat{\beta}}_{B}+\hat{\omega}_0\hat{J}^{\beta}_{A}\hat{J}^{\hat{\beta}}_{B}+\hat{\omega}_0\hat{J}^{\hat{\beta}}_{A}J^{\beta}_{B}-\hat{\omega}_0\hat{J}^{\hat{\beta}}_{A}\hat{K}_{B}\right),
\end{align}
where
\begin{align}
    &\frac{\left\langle\pi \mu_{-}\right\rangle\left\langle\pi \mu_{+}\right\rangle}{\left\langle\pi \alpha_{+}\right\rangle\left\langle\pi \alpha_{-}\right\rangle}=\hat{\omega}_{0}+\hat{\omega}_{1}\langle\pi\hat{ \beta}\rangle+\mathcal{O}(\langle\pi \hat{\beta}\rangle^{2})\notag\\
    &\hat{K}_{A}=-\frac{\kappa_{A}}{\langle\hat{\beta}\mu_+\rangle}U_{\hat{\kappa}}^{(1)}+\frac{\hat{\kappa}_A}{\langle\hat{\beta}\mu_-\rangle}U_{\kappa}^{(1)}. \notag
\end{align}
Adding them up and dropping the vanishing terms, we get the Lagrangian density of the resulting 4D theory 
\begin{align}
    \label{c.24}
    \mathcal{L}_{4D}=\varepsilon^{AB}\operatorname{tr}\left(\omega_0 J^{\hat{\beta}}_{A} J^{\beta}_{B}+\omega_0 J^{\beta}_{A} \hat{J}^{\hat{\beta}}_{B}-\omega_0 J^{\beta}_{A}K_{B}-\hat{\omega}_0\hat{J}^{\beta}_{A}\hat{J}^{\hat{\beta}}_{B}-\hat{\omega}_0\hat{J}^{\hat{\beta}}_{A}J^{\beta}_{B}+\hat{\omega}_0\hat{J}^{\hat{\beta}}_{A}\hat{K}_{B}\right),
\end{align}
with
\begin{align}
    \label{c.38}
    \omega_0=\frac{\langle\beta\mu_+\rangle\langle\beta\mu_-\rangle}{\langle\beta\alpha_+\rangle\langle\beta\alpha_-\rangle}, \qquad
    \hat{\omega}_0=\frac{\langle\hat{\beta}\mu_+\rangle\langle\hat{\beta}\mu_-\rangle}{\langle\hat{\beta}\alpha_+\rangle\langle\hat{\beta}\alpha_-\rangle}.
\end{align}
This Lagrangian density is invariant under the $\mathbb{Z}_2$ symmetry\footnote{The $\mathbb{Z}_2$ operation on $\beta$ coincides with the complex conjugation of the spinors.} 
\bea \label{Z2}
\beta \rightarrow \hat{\beta},\quad \hat{\beta} \rightarrow -\beta; \quad J\leftrightarrow \hat{J}.
\eea 
To get the 4D analogue of usual Yang-Baxter deformation one may impose the simplest involution relation $J=\hat{J}$. The Lagrangian density then reduces to 
\begin{align}
    \label{c.25}
    \mathcal{L}_{4D}'=\varepsilon^{AB}\operatorname{tr}\left(\hat{\omega}_0 J^{\hat{\beta}}_{A}\hat{K}_{B}-\omega_0 J^{\beta}_{A}K_{B}\right),
\end{align}
in addition to the topological terms like \eqref{WZ} from the two double poles
\bea 
\frac{1}{3}\int_{\mathbb{E}^4\times[0,1]} \(k_1\mu_{\beta,\beta}+k_2\mu_{\hat{\beta},\hat{\beta}}+k_3\mu_{\beta,\hat{\beta}}\)\wedge \on{tr}(\bar{\mathcal{J}}^3)
\eea 
where
\bea 
&&k_1=\omega_0\(\frac{\la\mu_-\alpha_-\ra}{\la\beta\alpha_-\ra \la\beta\mu_-\ra}+\frac{\la\mu_+\alpha_+\ra}{\la\beta\alpha_+\ra \la\beta\mu_+\ra}\),\notag\\
&& k_2=\hat{\omega}_0\(\frac{\la\mu_-\alpha_-\ra}{\la\hat{\beta}\alpha_-\ra \la\hat{\beta}\mu_-\ra}+\frac{\la\mu_+\alpha_+\ra}{\la\hat{\beta}\alpha_+\ra \la\hat{\beta}\mu_+\ra}\),\notag\\
&&k_3=2(\hat{\omega}_0-\omega_0).\notag
\eea

\subsection{Symmetry reduction to 2D theory}
Now we are ready to do the symmetry reduction. Imposing the symmetry \eqref{1.62} and introducing \eqref{add:1.23}, we can find
\begin{align}
    \label{1.63}
    \kappa^A\pi^{A'}\partial_{AA'}\sigma=&\kappa^A(\langle\pi\hat{\mu}\rangle\mu^{A'}-\langle\pi\mu\rangle\hat{\mu}^{A'})\partial_{AA'}\sigma=-\langle\pi\mu\rangle\partial_{w}\sigma,\\
    \label{1.64}
    \hat{\kappa}^A\pi^{A'}\partial_{AA'}\sigma=&\hat{\kappa}^A(\langle\pi\hat{\mu}\rangle\mu^{A'}-\langle\pi\mu\rangle\hat{\mu}^{A'})\partial_{AA'}\sigma=-\langle\pi\hat{\mu}\rangle\partial_{\bar{w}}\sigma.
\end{align}
Thus the right hand side of \eqref{c.24} becomes 
\begin{align}
    \label{c.42}
    \operatorname{tr} &\left[\frac{\langle\beta\mu_+\rangle\langle\beta\mu_-\rangle}{\langle\beta\alpha_+\rangle\langle\beta\alpha_-\rangle}\left(\langle\beta\mu\rangle\langle\hat{\beta}\hat{\mu}\rangle\partial_{w}\sigma\sigma^{-1}\partial_{\bar{w}}\hat{\sigma}\hat{\sigma}^{-1}-\langle\beta\hat{\mu}\rangle\langle\hat{\beta}\mu\rangle\partial_w\hat{\sigma}\hat{\sigma}^{-1}\partial_{\bar{w}}\sigma\sigma^{-1}-\partial_{w}\sigma\sigma^{-1}\partial_{\bar{w}}\sigma\sigma^{-1}\right)\right.\notag\\
    &+\frac{\langle\hat{\beta}\mu_+\rangle\langle\hat{\beta}\mu_-\rangle}{\langle\hat{\beta}\alpha_+\rangle\langle\hat{\beta}\alpha_-\rangle}\left(-\langle\hat{\beta}\mu\rangle\langle\beta\hat{\mu}\rangle\partial_{w}\hat{\sigma}\hat{\sigma}^{-1}\partial_{\bar{w}}\sigma\sigma^{-1}+\langle\hat{\beta}\hat{\mu}\rangle\langle\beta\mu\rangle\partial\sigma\sigma^{-1}\partial_{\bar{w}}\hat{\sigma}\hat{\sigma}^{-1}-\partial_{w}\hat{\sigma}\hat{\sigma}^{-1}\partial_{\bar{w}}\hat{\sigma}\hat{\sigma}^{-1}\right)\notag\\
    &\left.-\omega_0\frac{\langle\beta\hat{\mu}\rangle}{\langle\beta\mu_-\rangle}j_{\bar{w}}{U'}_{\kappa}^{(1)}+\omega_0\frac{\langle\beta\mu\rangle}{\langle\beta\mu_+\rangle}j_w {U'}_{\hat{\kappa}}^{(1)}+\hat{\omega}_0\frac{\langle\hat{\beta}\hat{\mu}\rangle}{\langle\hat{\beta}\mu_-\rangle}\hat{j}_{\bar{w}}{U'}_{\kappa}^{(1)}-\hat{\omega}_0\frac{\langle\hat{\beta}\mu\rangle}{\langle\hat{\beta}\mu_+\rangle}\hat{j}_w {U'}_{\hat{\kappa}}^{(1)}\right]
\end{align}
where $U'^{(1)}_{\kappa,\hat{\kappa}}$ are given by the same expressions as  \eqref{c.13} and \eqref{c.14} but with the symmetry reduced version of $U_{\kappa,\hat{\kappa}}, \hat{U}_{\kappa,\hat{\kappa}}$
\begin{align}
    \label{c.43}
    \hat{U}_{\kappa}=-\langle\beta\mu\rangle j_w,\quad
    \hat{U}_{\hat{\kappa}}=-\langle\beta\hat{\mu}\rangle j_{\bar{w}},\quad
    U_{\kappa}=\langle\hat{\beta}\mu\rangle\hat{j}_{w},\quad
    U_{\hat{\kappa}}=\langle\hat{\beta}\hat{\mu}\rangle \hat{j}_{\bar{w}}. 
\end{align}
On the other hand, contracting both sides of 
\be
\mathrm{d}^{2} x^{A^{\prime} B^{\prime}} \alpha_{+A^{\prime}} \alpha_{+B^{\prime}} \wedge \mathrm{d} x^{C C^{\prime}} \wedge \mathrm{d} x^{D D^{\prime}}=-2\mathrm{vol}_{4} \varepsilon^{C D}\alpha_+^{C^{\prime}} \alpha_+^{D^{\prime}}\notag
\ee
 with $\kappa_C\hat{\kappa}_{D}\mu_{C'}\hat{\mu}_{D'}$
 and acting $\iota_{\chi\wedge\bar{\chi}}$ on both sides of the resulting equation, we have
\begin{align}
    \label{1.67}
    -\iota_{\chi\wedge\bar{\chi}}\operatorname{vol}_4 =\mathrm{d}w\wedge\mathrm{d}\bar{w}. 
\end{align}
Therefore the resulting expression for the symmetry reduced 2D version of \eqref{c.24} is
\begin{align}
    \label{c.44}
    \mathcal{L}_{2D}=\operatorname{tr} &\left[\frac{\langle\beta\mu_+\rangle\langle\beta\mu_-\rangle}{\langle\beta\alpha_+\rangle\langle\beta\alpha_-\rangle}\left(-\langle\beta\mu\rangle\langle\hat{\beta}\hat{\mu}\rangle j_w \hat{j}_{\bar{w}}+\langle\beta\hat{\mu}\rangle\langle\hat{\beta}\mu\rangle\hat{j}_{w}j_{\bar{w}}+j_{w} j_{\bar{w}}\right)\right.\notag\\
    &+\frac{\langle\hat{\beta}\mu_+\rangle\langle\hat{\beta}\mu_-\rangle}{\langle\hat{\beta}\alpha_+\rangle\langle\hat{\beta}\alpha_-\rangle}\left(+\langle\hat{\beta}\mu\rangle\langle\beta\hat{\mu}\rangle\hat{j}_{w}j_{\bar{w}}-\langle\hat{\beta}\hat{\mu}\rangle\langle\beta\mu\rangle j_{w}\hat{j}_{\bar{w}}+\hat{j}_w\hat{j}_{\bar{w}}\right)\notag\\
    &\left.+\omega_0\frac{\langle\beta\hat{\mu}\rangle}{\langle\beta\mu_-\rangle}j_{\bar{w}}{U'}_{\kappa}^{(1)}-\omega_0\frac{\langle\beta\mu\rangle}{\langle\beta\mu_+\rangle}j_w {U'}_{\hat{\kappa}}^{(1)}-\hat{\omega}_0\frac{\langle\hat{\beta}\hat{\mu}\rangle}{\langle\hat{\beta}\mu_-\rangle}\hat{j}_{\bar{w}}{U'}_{\kappa}^{(1)}+\hat{\omega}_0\frac{\langle\hat{\beta}\mu\rangle}{\langle\hat{\beta}\mu_+\rangle}\hat{j}_w {U'}_{\hat{\kappa}}^{(1)}\right],
\end{align}
which also enjoys the $\mathbb{Z}_2$ symmetry \eqref{Z2}. Setting $j=\hat{j}$ we reproduce the standard action of the Yang-Baxter deformation 
\begin{align}
    \label{c.45}
    \mathcal{L}'_{2D}&=\operatorname{tr}\left(\omega_0\frac{\langle\beta\hat{\mu}\rangle}{\langle\beta\mu_-\rangle}j_{\bar{w}}{U'}_{\kappa}^{(1)}-\omega_0\frac{\langle\beta\mu\rangle}{\langle\beta\mu_+\rangle}j_w {U'}_{\hat{\kappa}}^{(1)}-\hat{\omega}_0\frac{\langle\hat{\beta}\hat{\mu}\rangle}{\langle\hat{\beta}\mu_-\rangle}\hat{j}_{\bar{w}}{U'}_{\kappa}^{(1)}+\hat{\omega}_0\frac{\langle\hat{\beta}\mu\rangle}{\langle\hat{\beta}\mu_+\rangle}\hat{j}_w {U'}_{\hat{\kappa}}^{(1)}\right) \notag\\
    &=\mathcal{N}\operatorname{tr}\left(j_w\frac{1}{\eta R-1}j_{\bar{w}}\right)
\end{align}
up to a overall factor
\begin{align}
    \label{x.1}
    \mathcal{N}=i(\eta'-i)&\left(\frac{\langle\beta\mu\rangle\langle\beta\mu_-\rangle\langle\alpha_+\mu_+\rangle\langle\alpha_+\hat{\mu}\rangle}{\langle\beta\alpha_+\rangle\langle\beta\alpha_-\rangle\langle\alpha_+\hat{\beta}\rangle\langle\alpha_+\beta\rangle}-\frac{\langle\beta\hat{\mu}\rangle\langle\beta\mu_+\rangle\langle\alpha_-\mu_-\rangle\langle\alpha_-\mu\rangle}{\langle\beta\alpha_+\rangle\langle\beta\alpha_-\rangle\langle\alpha_-\hat{\beta}\rangle\langle\alpha_-\beta\rangle}\right.\notag\\
    &\left.-\frac{\langle\hat{\beta}\mu\rangle\langle\hat{\beta}\mu_-\rangle\langle\alpha_+\mu_+\rangle\langle\alpha_+\hat{\mu}\rangle}{\langle\hat{\beta}\alpha_+\rangle\langle\hat{\beta}\alpha_-\rangle\langle\alpha_+\hat{\beta}\rangle\langle\alpha_+\beta\rangle}+\frac{\langle\hat{\beta}\hat{\mu}\rangle\langle\hat{\beta}\mu_+\rangle\langle\alpha_-\mu_-\rangle\langle\alpha_-\mu\rangle}{\langle\hat{\beta}\alpha_+\rangle\langle\hat{\beta}\alpha_-\rangle\langle\alpha_-\hat{\beta}\rangle\langle\alpha_-\beta\rangle}\right). 
\end{align}

\subsection{Symmetry reduction of 6DhCS to 4D Chern-Simons theory}
In this subsection we discuss the left route in the diagram in Fig. 4. Since the symmetry group of reduction is the same as before, we can use the general expression \eqref{equ:1.20} and \eqref{equ:1.23} directly to obtain the reduced 4D CS theory. Our choice of $(3,0)$-form $\Omega$ leads to the 1-form 
\begin{align}
    \label{1.24}
    \omega=\frac{\langle\mathrm{d}\pi \pi\rangle\langle\pi\mu\rangle\langle\pi\hat{\mu}\rangle\langle\pi\mu_{+}\rangle\langle\pi\mu_{-}\rangle}{\langle\pi\alpha_+\rangle\langle\pi\alpha_-\rangle\langle\pi\beta\rangle^2\langle\pi\hat{\beta}\rangle^2}. 
\end{align}
with the condition \eqref{c.1}. If we use the \textit{matching condition} \eqref{add:1.25} and \eqref{add:1.26} directly then the boundary condition of 4DCS would be
\begin{align}
    \label{c.3}
    \langle\alpha_{+}\mu\rangle(R-i)\bar{A}_w|_{\pi=\alpha_{+}}=\langle\alpha_-\mu\rangle(R+i)\bar{A}_w|_{\pi=\alpha_-},\\
    \langle\alpha_{+}\hat{\mu}\rangle(R-i)\bar{A}_{\bar{w}}|_{\pi=\alpha_{+}}=\langle\alpha_-\hat{\mu}\rangle(R+i)\bar{A}_{\bar{w}}|_{\pi=\alpha_-},\\
    \label{c.4}
    \left.\bar{A}_{w,\bar{w}}\right|_{\pi=\beta}=0, \quad \left. \bar{A}_{w,\bar{w}}\right|_{\pi=\hat{\beta}}=0.
\end{align}
We encounter the same problem that we revealed in the example of $\lambda$--deformation.
To proceed we should not insist on the exact matching of the boundary conditions in 6DhCS and 4DCS, and that is why we used dotted line in the diagram in Fig. 4.
Therefore we just start from \eqref{1.24} and introduce a new condition
\begin{align}
    \label{c.47}
    \frac{\langle\alpha_+\mu\rangle\langle\alpha_+\hat{\mu}\rangle\langle\alpha_{+}\mu_{+}\rangle\langle\alpha_{+}\mu_{-}\rangle}{\langle\alpha_{+}\beta\rangle^2\langle\alpha_{+}\hat{\beta}\rangle^2}=\frac{\langle\alpha_-\mu\rangle\langle\alpha_-\hat{\mu}\rangle\langle\alpha_{-}\mu_{+}\rangle\langle\alpha_{-}\mu_{-}\rangle}{\langle\alpha_{-}\beta\rangle^2\langle\alpha_{-}\hat{\beta}\rangle^2},
\end{align}
to ensure the residues at the two simple poles are opposite. Correspondingly the boundary conditions should be chosen to be
\begin{align}
    \label{1.25}
    \left.(R-i)\bar{A}_{w,\bar{w}}\right|_{\pi=\alpha_+}=\left.(R+i)\bar{A}_{w,\bar{w}}\right|_{\pi=\alpha_-},\\
    \label{1.26}
    \left.\bar{A}_{w,\bar{w}}\right|_{\pi=\beta}=0, \quad \left. \bar{A}_{w,\bar{w}}\right|_{\pi=\hat{\beta}}=0. 
\end{align}
In particular the boundary condition is not equivalent to \eqref{c.6} and \eqref{c.7} which one can check easily using \eqref{equ:1.20}. Nevertheless, since the gauge symmetry is same, we can still use a similar argument above \eqref{Fix} to remove the gauge freedoms by fixing
\begin{align}
\label{1.28}
    \tilde{\sigma}|_{\beta}\equiv\sigma,\qquad \tilde{\sigma}|_{\hat{\beta}}\equiv\hat{\sigma}, \qquad \tilde{\sigma}|_{\alpha_+}=\operatorname{id},\qquad \tilde{\sigma}|_{\alpha_-}=\operatorname{id}.
\end{align}
The equations \eqref{c.7}, \eqref{c.8} and \eqref{equ:1.20} imply that the 2D Lax connection has to be in the form
\begin{align}
    \label{c.26}
    \mathcal{L}_{w}&=\frac{\langle\pi\beta\rangle\langle\pi\hat{\beta}\rangle}{\langle\pi\mu\rangle\langle\pi\mu_-\rangle}V^{(1)}_{w}+\frac{\langle\pi\beta\rangle}{\langle\pi\mu\rangle} V_{w}+\frac{\langle\pi\hat{\beta}\rangle}{\langle\pi\mu\rangle} \bar{V}_{w},\\
    \label{c.27}
    \mathcal{L}_{\bar{w}}&=\frac{\langle\pi\beta\rangle\langle\pi\hat{\beta}\rangle}{\langle\pi\hat{\mu}\rangle\langle\pi\mu_+\rangle}V^{(1)}_{\bar{w}}+\frac{\langle\pi\beta\rangle}{\langle\pi\hat{\mu}\rangle} V_{\bar{w}}+\frac{\langle\pi\hat{\beta}\rangle}{\langle\pi\hat{\mu}\rangle} \bar{V}_{\bar{w}},
\end{align}
where $V^{(1)}_{w,\bar{w}}, V_{w,\bar{w}}$ and $\bar{V}_{w,\bar{w}}$ do not depend on the coordinates $\pi$ and $\hat{\pi}$ of $\mathbb{CP}^1$.  
Substituting the ansatz \eqref{c.26} and \eqref{c.27} into the boundary conditions \eqref{1.25} and \eqref{1.26} gives
\begin{align}
    \label{c.28}
    \bar{V}_{w}&=-\langle\beta\mu\rangle \partial_{w}\sigma \sigma^{-1} \equiv\langle\beta\mu\rangle j_{w},\qquad \bar{V}_{\bar{w}}=-\langle\beta\hat{\mu}\rangle \partial_{\bar{w}}\sigma \sigma^{-1} \equiv\langle\beta\hat{\mu}\rangle j_{\bar{w}},\\
    \label{c.29}
    V_{w}&=\langle\hat{\beta}\mu\rangle \partial_{w}\hat{\sigma} \hat{\sigma}^{-1} \equiv-\langle\hat{\beta}\mu\rangle \hat{j}_{w},\qquad V_{\bar{w}}=\langle\hat{\beta}\hat{\mu}\rangle \partial_{\bar{w}}\hat{\sigma} \hat{\sigma}^{-1} \equiv-\langle\hat{\beta}\hat{\mu}\rangle \hat{j}_{\bar{w}},
\end{align}
and 
\begin{align}
    \label{c.30}
    &(R-i)\left(\frac{\langle\alpha_+\beta\rangle\langle\alpha_+\hat{\beta}\rangle}{\langle\alpha_+\mu\rangle\langle\alpha_+\mu_-\rangle}V^{(1)}_{w}+\frac{\langle\alpha_+\beta\rangle}{\langle\alpha_+\mu\rangle} V_{w}+\frac{\langle\alpha_+\hat{\beta}\rangle}{\langle\alpha_+\mu\rangle} \bar{V}_{w}\right)\notag\\
    &\hspace{3ex}=(R+i)\left(\frac{\langle\alpha_-\beta\rangle\langle\alpha_-\hat{\beta}\rangle}{\langle\alpha_-\mu\rangle\langle\alpha_-\mu_-\rangle}V^{(1)}_{w}+\frac{\langle\alpha_-\beta\rangle}{\langle\alpha_-\mu\rangle} V_{w}+\frac{\langle\alpha_-\hat{\beta}\rangle}{\langle\alpha_-\mu\rangle} \bar{V}_{w}\right),\\
    \label{c.31}
    &(R-i)\left(\frac{\langle\alpha_+\beta\rangle\langle\alpha_+\hat{\beta}\rangle}{\langle\alpha_+\hat{\mu}\rangle\langle\alpha_+\mu_+\rangle}V^{(1)}_{\bar{w}}+\frac{\langle\alpha_+\beta\rangle}{\langle\alpha_+\hat{\mu}\rangle} V_{\bar{w}}+\frac{\langle\alpha_+\hat{\beta}\rangle}{\langle\alpha_+\hat{\mu}\rangle} \bar{V}_{\bar{w}}\right)\notag\\
    &\hspace{3ex}=(R+i)\left(\frac{\langle\alpha_-\beta\rangle\langle\alpha_-\hat{\beta}\rangle}{\langle\alpha_-\hat{\mu}\rangle\langle\alpha_-\mu_+\rangle}V^{(1)}_{\bar{w}}+\frac{\langle\alpha_-\beta\rangle}{\langle\alpha_-\hat{\mu}\rangle} V_{\bar{w}}+\frac{\langle\alpha_-\hat{\beta}\rangle}{\langle\alpha_-\hat{\mu}\rangle} \bar{V}_{\bar{w}}\right). 
\end{align}
The solution of our ansatz is
\begin{align}
    \label{c.32}
    V_{w}^{(1)}&=-(\eta'-i) \frac{\langle\alpha_{-} \mu_{-}\rangle}{2\langle\alpha_{-} \hat{\beta}\rangle\langle\alpha_{-} \beta\rangle} \frac{\left(\gamma^{w}_{-} R+i \gamma^{w}_{+}\right) V_{w}+\left(\hat{\gamma}^{w}_{-} R+i \hat{\gamma}^{w}_{+}\right) \bar{V}_{w}}{\eta R+1},\\
    \label{c.33}
    V_{\bar{w}}^{(1)}&=(\eta'-i)\frac{\langle\alpha_{+} \mu_{+}\rangle}{2\langle\alpha_{+} \hat{\beta}\rangle\langle\alpha_{+} \beta\rangle}\frac{\left(\gamma^{\bar{w}}_{-} R+i \gamma^{\bar{w}}_{+}\right) V_{\bar{w}}+\left(\hat{\gamma}^{\bar{w}}_{-} R+i \hat{\gamma}^{\bar{w}}_{+}\right) \bar{V}_{\bar{w}}}{\eta R-1},
\end{align}
where we have introduced the following quantities
\begin{align}
    \gamma^{w}_{\pm}&=\langle\alpha_-\beta\rangle\pm\langle\alpha_+\beta\rangle\frac{\langle\alpha_-\mu\rangle}{\langle\alpha_+\mu\rangle},\quad \hat{\gamma}^{w}_{\pm}=\langle\alpha_-\hat{\beta}\rangle\pm\langle\alpha_+\hat{\beta}\rangle\frac{\langle\alpha_-\mu\rangle}{\langle\alpha_+\mu\rangle}, \notag\\
    \gamma^{\bar{w}}_{\pm}&=\langle\alpha_-\beta\rangle\frac{\langle\alpha_+\hat{\mu}\rangle}{\langle\alpha_-\hat{\mu}\rangle}\pm\langle\alpha_+\beta\rangle,\quad \hat{\gamma}^{\bar{w}}_{\pm}=\langle\alpha_-\hat{\beta}\rangle\frac{\langle\alpha_+\hat{\mu}\rangle}{\langle\alpha_-\hat{\mu}\rangle}\pm\langle\alpha_+\hat{\beta}\rangle,\notag\\
    \eta'&=-i \frac{1-\frac{\left\langle\alpha_{-} \beta\right\rangle\left\langle\alpha_{-} \hat{\beta}\right\rangle\left\langle\alpha_{+} \mu_{+}\right\rangle\langle\alpha_+\hat{\mu}\rangle}{\left\langle\alpha_{+} \beta\right\rangle\left\langle\alpha_{+} \hat{\beta}\right\rangle\left\langle\alpha_{-} \mu_{+}\right\rangle\langle\alpha_-\hat{\mu}\rangle}}{1+\frac{\left\langle\alpha_{-} \beta\right\rangle\left\langle\alpha_{-} \hat{\beta}\right\rangle\left\langle\alpha_{+} \mu_{+}\right\rangle\langle\alpha_+\hat{\mu}\rangle}{\left\langle\alpha_{+} \beta\right\rangle\left\langle\alpha_{+} \hat{\beta}\right\rangle\left\langle\alpha_{-} \mu_{+}\right\rangle\alpha_-\hat{\mu}\rangle}}=-i \frac{1-\frac{\left\langle\alpha_{+} \beta\right\rangle\left\langle\alpha_{+} \hat{\beta}\right\rangle\left\langle\alpha_{-} \mu_{-}\right\rangle\langle\alpha_-\mu\rangle}{\left\langle\alpha_{-} \beta\right\rangle\left\langle\alpha_{-} \hat{\beta}\right\rangle\left\langle\alpha_{+} \mu_{-}\right\rangle\langle\alpha_+\mu\rangle}}{1+\frac{\left\langle\alpha_{+} \beta\right\rangle\left\langle\alpha_{+} \hat{\beta}\right\rangle\left\langle\alpha_{-} \mu_{-}\right\rangle\alpha_-\mu\rangle}{\left\langle\alpha_{-} \beta\right\rangle\left\langle\alpha_{-} \hat{\beta}\right\rangle\left\langle\alpha_{+} \mu_{-}\right\rangle\langle\alpha_+\mu\rangle}}.\notag
\end{align}
Finally substituting the 2D Lax connection into the 4DCS action, we end up with 
\begin{align}
    \label{1.39}
    S_{\omega}\left[\bar{A}\right]=&\frac{1}{2 \pi i} \int_{\mathbb{E}^{2} \times \mathbb{C} \mathbb{P}^{1}} \omega \wedge \operatorname{phCS}\left(\bar{A}\right)\notag\\
    =&\frac{1}{2 \pi i}\sum_{z\in\mathfrak{p}} \lim _{\epsilon \rightarrow 0} \oint_{S_{z, \epsilon}^{1}}\left[\frac{\langle\mathrm{d}\pi \pi\rangle\langle\pi\mu\rangle\langle\pi\hat{\mu}\rangle\langle\pi\mu_{+}\rangle\langle\pi\mu_{-}\rangle}{\langle\pi\alpha_+\rangle\langle\pi\alpha_-\rangle\langle\pi\beta\rangle^2\langle\pi\hat{\beta}\rangle^2} \int_{\mathbb{E}^{2}}  \operatorname{tr}(j \wedge \mathcal{L})\right]\notag\\
    &+\frac{1}{6\pi i}\int_{\mathbb{E}^{2} \times \mathbb{C} \mathbb{P}^{1}}\frac{\langle\mathrm{d}\pi \pi\rangle\langle\pi\mu\rangle\langle\pi\hat{\mu}\rangle\langle\pi\mu_{+}\rangle\langle\pi\mu_{-}\rangle}{\langle\pi\alpha_+\rangle\langle\pi\alpha_-\rangle\langle\pi\beta\rangle^2\langle\pi\hat{\beta}\rangle^2}\operatorname{tr}(\tilde{j}^3),
\end{align}
where $\tilde{j}\equiv-\mathrm{d}'\tilde{\sigma}\tilde{\sigma}^{-1}$. There are also two  contributions to the action coming from the terms linear in $\langle\pi\beta\rangle$ and the terms linear in $\langle\pi\hat{\beta}\rangle$ in the expansion of 
\begin{align}
    \frac{\langle\pi\mu\rangle\langle\pi\hat{\mu}\rangle\langle\pi\mu_{+}\rangle\langle\pi\mu_{-}\rangle}{\langle\pi\alpha_+\rangle\langle\pi\alpha_-\rangle}\operatorname{tr}(j\wedge \mathcal{L}). \notag
\end{align}
From those terms proportional to $\langle\pi\beta\rangle$, the contribution to Lagrangian is
\begin{align}
    &\frac{1}{2\pi i} \operatorname{tr}\left(\omega'_1 j_{w}\frac{1}{\langle\beta\hat{\mu}\rangle} \langle\beta\hat{\mu}\rangle j_{\bar{w}}-\omega'_0 j_{w}\frac{1}{\langle\beta\hat{\mu}\rangle} \langle\hat{\beta}\hat{\mu}\rangle \hat{j}_{\bar{w}}  +\omega'_0 j_{w}\frac{\langle\hat{\beta}\hat{\mu}\rangle}{\langle\beta\hat{\mu}\rangle^2}\langle\beta\hat{\mu}\rangle j_{\bar{w}}+\omega'_0 j_{w} \frac{1}{\langle\beta\hat{\mu}\rangle\langle\beta\mu_+\rangle} V_{\bar{w}}^{(1)}\right)\notag\\
    -&\frac{1}{2\pi i}\operatorname{tr}\left(\omega'_1 \frac{1}{\langle\beta\mu\rangle}\langle\beta\mu\rangle j_{w} j_{\bar{w}}-\omega'_0 \frac{1}{\langle\beta\mu\rangle}\langle\hat{\beta}\mu\rangle \hat{j}_{w} j_{\bar{w}}+\omega'_0 \frac{\langle\hat{\beta}\mu\rangle}{\langle\beta\mu\rangle^2}\langle\beta\mu\rangle j_{w} j_{\bar{w}}+\omega'_0 \frac{1}{\langle\beta\mu\rangle\langle\beta\mu_-\rangle}V_{w}^{(1)} j_{\bar{w}}\right)\notag
\end{align}
where
\begin{align}
    \frac{\langle\pi\mu\rangle\langle\pi\hat{\mu}\rangle\langle\pi\mu_{+}\rangle\langle\pi\mu_{-}\rangle}{\langle\pi\alpha_+\rangle\langle\pi\alpha_-\rangle}=\omega'_0+\omega'_1\langle\pi\beta\rangle+\mathcal{O}(\langle\pi\beta\rangle^2),\notag
\end{align}
and the contribution from the term proportional to $\langle\pi\hat{\beta}\rangle$ is
\begin{align}
    &\frac{1}{2\pi i}\operatorname{tr}\left(\hat{\omega}'_{1}\hat{j}_{w}\frac{1}{\langle\hat{\beta}\hat{\mu}\rangle}\langle\hat{\beta}\hat{\mu}\rangle \hat{j}_{\bar{w}}+\hat{\omega}'_0 \hat{j}_{w}\frac{1}{\langle\hat{\beta}\hat{\mu}\rangle}\langle\beta\hat{\mu}\rangle j_{\bar{w}}-\hat{\omega}'_0 \hat{j}_{w} \frac{\langle\beta\hat{\mu}\rangle}{\langle\hat{\beta}\hat{\mu}\rangle^2} \langle\hat{\beta}\hat{\mu}\rangle \hat{j}_{\bar{w}}-\hat{\omega}'_0 \hat{j}_{w}\frac{1}{\langle\hat{\beta}\hat{\mu}\rangle\langle\hat{\beta}\mu_+\rangle} V_{\bar{w}}^{(1)}\right)\notag\\
    -&\frac{1}{2\pi i}\operatorname{tr}\left(\hat{\omega}'_1 \frac{1}{\langle\hat{\beta}\mu\rangle}\langle\hat{\beta}\mu\rangle \hat{j}_{w} \hat{j}_{\bar{w}}+\hat{\omega}'_0 \frac{1}{\langle\hat{\beta}\mu\rangle}\langle\beta\mu\rangle j_{w} \hat{j}_{\bar{w}}-\hat{\omega}'_0 \frac{\langle\beta\mu\rangle}{\langle\hat{\beta}\mu\rangle^2}\langle\hat{\beta}\mu\rangle \hat{j}_{w} \hat{j}_{\bar{w}} -\hat{\omega}'_0\frac{1}{\langle\hat{\beta}\mu\rangle\langle\hat{\beta}\mu_-\rangle}V^{(1)}_w \hat{j}_{\bar{w}} \right)\notag
\end{align}
where
\begin{align}
    \frac{\langle\pi\mu\rangle\langle\pi\hat{\mu}\rangle\langle\pi\mu_{+}\rangle\langle\pi\mu_{-}\rangle}{\langle\pi\alpha_+\rangle\langle\pi\alpha_-\rangle}=\hat{\omega}'_0+\hat{\omega}'_1\langle\pi\hat{\beta}\rangle+\mathcal{O}(\langle\pi\hat{\beta}\rangle^2). \notag
\end{align}
Combing these two contributions together, we get the Lagrangian density of the 2D integrable theory
\begin{align}
    \label{c.41}
    \tilde{\mathcal{L}}_{2D}=\operatorname{tr}&\left[\frac{\langle\beta\mu_+\rangle\langle\beta\mu_-\rangle}{\langle\beta\alpha_+\rangle\langle\beta\alpha_-\rangle}\left(-\langle\beta\mu\rangle\langle\hat{\beta}\hat{\mu}\rangle j_{w}\hat{j}_{\bar{w}}+\langle\hat{\beta}\mu\rangle\langle\beta\hat{\mu}\rangle\hat{j}_{w}j_{\bar{w}}+j_w j_{\bar{w}}\right)\right.\notag\\
    &+\frac{\langle\hat{\beta}\mu_+\rangle\langle\hat{\beta}\mu_-\rangle}{\langle\hat{\beta}\alpha_+\rangle\langle\hat{\beta}\alpha_-\rangle}\left(\langle\hat{\beta}\mu\rangle\langle\beta\hat{\mu}\rangle\hat{j}_{w}j_{\bar{w}}-\langle\beta\mu\rangle\langle\hat{\beta}\hat{\mu}\rangle j_{w}\hat{j}_{\bar{w}}+\hat{j}_w \hat{j}_{\bar{w}}\right)\notag\\
    &+\left.\frac{\omega'_0 j_w V_{\bar{w}}^{(1)}}{\langle\beta\hat{\mu}\rangle\langle\beta\mu_+\rangle}-\frac{\omega'_0 V_{w}^{(1)} j_{\bar{w}}}{\langle\beta\mu\rangle\langle\beta\mu_-\rangle}-\frac{\hat{\omega}'_0 \hat{j}_w V_{\bar{w}}^{(1)}}{\langle\hat{\beta}\hat{\mu}\rangle\langle\hat{\beta}\mu_+\rangle}+\frac{\hat{\omega}'_0V_w^{(1)}\hat{j}_{\bar{w}}}{\langle\hat{\beta}\mu\rangle\langle\hat{\beta}\mu_-\rangle}\right],
\end{align}
with
\begin{align}
    \label{c.40}
    \omega'_0=\frac{\langle\beta\mu\rangle\langle\beta\hat{\mu}\rangle\langle\beta\mu_{+}\rangle\langle\beta\mu_{-}\rangle}{\langle\beta\alpha_+\rangle\langle\beta\alpha_-\rangle}, \qquad \hat{\omega}'_0=\frac{\langle\hat{\beta}\mu\rangle\langle\hat{\beta}\hat{\mu}\rangle\langle\hat{\beta}\mu_{+}\rangle\langle\hat{\beta}\mu_{-}\rangle}{\langle\hat{\beta}\alpha_+\rangle\langle\hat{\beta}\alpha_-\rangle}. 
\end{align}
Note that this is almost same as \eqref{c.44} but with only ${U'}^{(1)}_{\kappa}$, ${U'}^{(1)}_{\hat{\kappa}}$ getting replaced by $-V^{(1)}_w$ and $-V^{(1)}_{\bar{w}}$, so this 2D theory is still invariant under the $\mathbb{Z}_2$ symmetry \eqref{Z2} . Imposing the involution $j=\hat{j}$ up to a overall factor we reproduce the familiar action of the Yang-Baxter model 
\begin{align}
    \label{c.46}
    \tilde{\mathcal{L}}'_{2D}&=\operatorname{tr}\left(\frac{\omega'_0 j_w V_{\bar{w}}^{(1)}}{\langle\beta\hat{\mu}\rangle\langle\beta\mu_+\rangle}-\frac{\omega'_0 V_{w}^{(1)} j_{\bar{w}}}{\langle\beta\mu\rangle\langle\beta\mu_-\rangle}-\frac{\hat{\omega}'_0 \hat{j}_w V_{\bar{w}}^{(1)}}{\langle\hat{\beta}\hat{\mu}\rangle\langle\hat{\beta}\mu_+\rangle}+\frac{\hat{\omega}'_0V_w^{(1)}\hat{j}_{\bar{w}}}{\langle\hat{\beta}\mu\rangle\langle\hat{\beta}\mu_-\rangle}\right)\notag\\
    &=\mathcal{N}\operatorname{tr}\left(j_w\frac{1}{\eta R-1}j_{\bar{w}}\right). 
\end{align}
This  concludes the ``commutativity" of the diagram in Fig. 4.

\section{Conclusions}
In this work we  studied the relations between 6DhCS, 4DCS and 2D integrable theories for more general cases. We found that the diagram in Fig. \ref{Figure1} proposed in \cite{Skiner} is not always commutative due to the fact that the \textit{matching condition} between the 6D and 4D theory is not always compatible with the boundary conditions. More precisely, the ascending or descending of the boundary conditions in one theory to the boundary conditions in the other  could be problematic in the sense that the ascended or descended boundary condition can not remove the gauge freedoms and the resulting theory has gauge symmetry and is not the theory we want. Even if we discard the \textit{matching condition} and impose appropriate boundary conditions to remove the gauge freedoms, the resulting 2D models in two approaches are often different. For the $\lambda$--deformation, the 2D model read from 4D WZW model is either trivial or the deformation parameter is restricted to specific value. For the rational $\eta$--deformation, even though the 4D WZW-like model is integrable, its symmetry reduction to 2D is not integrable anymore.  Furthermore, we investigated two more cases that $\Omega$ are of more general forms including zeros, without insisting on the \textit{matching condition}. 
In particular, we managed to construct $\eta$--deformation in the trigonometric description from both two routes in the diagram in Fig. 4. What's more, we also obtained a coupled version of $\lambda$--deformation from both roads by considering higher-order pole in $\Omega$.  \par

It is well-known that the $\lambda$--deformation is related to the $\eta$--deformation through Poisson-Lie-T-duality and analytic continuation. If there is a trigonometric description for the $\lambda$--deformation, then it would be also possible to construct a 4D $\lambda$--deformation from 6DhCS. It will be also interesting to study 4D $\lambda$--deformation from the point of view of 4D WZW model directly following the original construction of $\lambda$--deformation in 2D \cite{Lambda}. In \cite{CWY,Fukushima:2020dcp,Costello:2020lpi,Tian:2020ryu,Tian:2020meg}, the coset models and supercoset models have been successfully constructed from 4DCS, then it is interesting to investigate their constructions from 6DhCS theory.

Another closely related model is the affine Gaudin model \cite{Gaudin} which also can be used to construct 2D integrable field theories systematically. The relation between the affine Gaudin model and 4DCS has been studied in \cite{Vicedo:2019dej}. It would be interesting to see how the affine Gaudin model fits in the diagram in Fig. \ref{Figure1}.

\section*{Acknowledgments}
JT and YJH would like to thank to thank Jue Hou, Reiko Liu and Jun-Bao Wu for useful discussion. BC and YJH were in part supported by NSFC Grant  No. 11735001. JT is also supported by the UCAS program of special research associate and by the internal funds of the KITS.

\vspace{1cm}

\bibliographystyle{JHEP}

\end{document}